%% file: main.tex
\documentclass[aps,pra,reprint,raggedbottom,superscriptaddress,amsmath,amssymb,longbibliography]{revtex4-2}
\usepackage[T1]{fontenc}
\usepackage[utf8]{inputenc}
\usepackage{lmodern}
\usepackage{amsthm}
\usepackage{graphicx}
\usepackage{needspace}
\usepackage{xcolor}
\definecolor{RefBlue}{RGB}{0,65,130}
\usepackage{hyperref}
\hypersetup{colorlinks=true,linkcolor=RefBlue,citecolor=RefBlue,urlcolor=RefBlue,pdfauthor={Yibin Huang, Muchun Yang, and D. L. Zhou},pdftitle={Minimal Projective Measurement Settings for Estimating Multiple Observables}}
\makeatletter
\g@addto@macro\normalsize{%
  \abovedisplayskip 4.5pt plus 2pt minus 2pt
  \belowdisplayskip \abovedisplayskip
  \abovedisplayshortskip 2.5pt plus 2pt minus 2pt
  \belowdisplayshortskip \abovedisplayshortskip
}
\makeatother
\newtheorem{theorem}{Theorem}
\renewcommand{\acknowledgmentsname}{Acknowledgements}

\begin{document}
\title{Minimal Projective Measurement Settings for Estimating Multiple Observables}
\author{Yibin Huang}
\affiliation{Institute of Physics, Beijing National Laboratory for Condensed
  Matter Physics,\\Chinese Academy of Sciences, Beijing 100190, China}
\affiliation{School of Physical Sciences, University of Chinese Academy of
  Sciences, Beijing 100049, China}

\author{Muchun Yang}
\email[]{yangmuchun21@mails.ucas.edu.cn}
\affiliation{Department of Physics, Tsinghua University, Beijing 100084, China}

\author{D. L. Zhou}
\email[]{zhoudl72@iphy.ac.cn}
\affiliation{Institute of Physics, Beijing National Laboratory for Condensed
  Matter Physics,\\Chinese Academy of Sciences, Beijing 100190, China}
\affiliation{School of Physical Sciences, University of Chinese Academy of
  Sciences, Beijing 100049, China}

\date{October 26, 2026}

\begin{abstract}
\input{abstract}
\end{abstract}
\maketitle

\section{Introduction}
\input{introduction}

\section{Main results}
\label{sec:main-results}
\input{main_results}

\section{Measurement framework and optimization}
\label{sec:framework}
\input{problem_formulation}
\input{witnesses}

\section{Numerical examples}
\label{sec:example}
\subsection{Collective spin moments of two qubits}
\label{sec:collective-moments}
\input{collective_example}
\subsection{Exact reconstruction on four qubits}
\label{sec:four-qubit}
\input{example}
\input{six_qubit_figure}
\subsection{Finite-depth measurements on six qubits}
\label{sec:six-qubit}
\input{six_qubit_example}

\section{Conclusion}
\label{sec:conclusion}
\input{conclusion}

\begin{acknowledgments}
\input{acknowledgements}
\end{acknowledgments}

\onecolumngrid
\clearpage
\twocolumngrid
\appendix
\section{Measurement protocols}
\label{app:protocols}
\input{measurement_protocols}
\input{six_qubit_protocols}
\input{baseline_convergence}
\input{initialization_checks}

\section{Proofs of the lower bounds}
\label{app:witness-proofs}
\input{structural_capacity_proof}
\input{witness_proofs}
\input{four_device_example}

\section{Error for approximate reconstruction}
\label{app:approximate-error}
\input{approximate_error}

\bibliography{references}
\end{document}

%% file: abstract.tex
We study the minimum number of orthogonal projective measurement settings needed to estimate multiple observables without bias for arbitrary input states. We derive structural bounds that go beyond dimension counting and combine them with explicit constructions to certify minimum setting counts for selected target spaces. A classical variational algorithm searches for shared measurement bases and optimizes their readout coefficients and sample allocation. For the collective first and second moments of two qubits, we certify that four projective settings are necessary and sufficient. For this task, a numerically optimized four-setting design achieves a Haar-averaged variance within approximately $0.0341\%$ of a lower bound valid for any number of projective settings in our fixed-allocation, unbiased linear estimation model. Numerical examples show that shared measurements can improve precision relative to the comparison protocols. For finite-depth circuits, we account for reconstruction bias and identify finite-sample regimes in which this advantage persists.

%% file: introduction.tex
Estimating the expectation values of multiple observables is a common task in quantum information. Commuting observables can share measurement settings \cite{Verteletskyi2020Clique,Yen2020Compatible,Gokhale2020MeasurementCost,Hamamura2020Entangled}, while basis rotations can reduce the number of groups \cite{Izmaylov2020Unitary,Zhao2020Reduction,Yen2021Cartan,Huggins2021BasisRotation}. Optimizing sample allocation and using overlapping groups can reduce statistical errors \cite{Crawford2021Sampling,Yen2023Deterministic,Wu2023Overlapped,Shlosberg2023Adaptive,Zhu2024ShotAssignment}. Related measurement schedules reconstruct many few-body reduced density matrices \cite{Cotler2020Overlapping,BonetMonroig2020PartialTomography}.

Shadow tomography estimates many properties without reconstructing the full state \cite{Aaronson2018ShadowTomography}. Classical shadows reuse randomized measurement data for many observables \cite{Huang2020Shadows,Elben2023Toolbox}. Nonuniform sampling and deterministic measurement choices can improve accuracy for known observables \cite{Hadfield2022LocallyBiased,Huang2021Derandomization}. Generalized and joint measurements allow further optimization \cite{GarciaPerez2021Learning,Nguyen2022Generalized,McNulty2023Joint,Acharya2021POVMShadows}. Shallow classical shadows relate circuit depth and noise to sampling costs \cite{Hu2023LocallyScrambled,Ippoliti2023OptimalDepth,Bertoni2024Shallow,Rozon2024Noise}. Optimizing classical postprocessing can reduce variance for fixed measurements \cite{DAriano2007Optimal,Fischer2024DualFrame,Malmi2024Enhanced}, and a small bias can reduce the mean squared error at finite sample sizes \cite{Cai2025Biased}. Quantum neural networks also learn measurement schemes for observables \cite{Wu2024Randomness,Yang2025Incompatible}.

Orthogonal projective measurements are a basic tool for estimating the expectation values of observables in quantum information experiments. Each measurement setting applies a fixed unitary transformation to the state, followed by an orthogonal projective measurement in the computational basis. When the observables commute pairwise, we can estimate all their expectation values from the same data by measuring in a common eigenbasis. In other words, we need only one measurement setting. If the observables do not all commute with one another, one such measurement setting is not enough. This leads us to ask: Given a set of observables, how many such measurement settings are sufficient to estimate all their expectation values? We aim to minimize the number of settings. We study both the number of settings and the precision we can achieve with a finite number of state copies.

Closely related work includes Cartan-based observable decompositions \cite{Yen2021Cartan}, target-adapted generalized measurements \cite{Nguyen2022Generalized}, and minimal-setting protocols for universal shadow estimation \cite{Yang2026MinimalSettings}. Here we seek the minimum number of orthogonal projective settings required to estimate observables in a fixed target space without bias. We allow arbitrary global basis rotations and require coverage only of the targets, rather than information completeness. Our structural lower bounds constrain every choice of measurement bases and every decomposition across them. Together with matching constructions, they certify exact minimum setting counts for selected tasks, beyond what dimension counting alone can establish.

%% file: main_results.tex
We use a classical variational algorithm to jointly optimize the measurement unitaries, estimator coefficients, and sample allocation. An exact decomposition gives an upper bound on the minimum number of settings.

We derive a structural lower bound and a witness based on alternating traces that rules out two settings (Theorems~\ref{thm:structural-capacity} and~\ref{thm:odd-trace-witness}). Matching these bounds with constructions determines the minimum number of settings for selected target spaces. For our $30$ four-qubit observables, exactly three settings suffice.

The structural bound also determines the minimum for a natural task: estimating the collective first and second moments of two qubits (Sec.~\ref{sec:collective-moments}). It excludes three arbitrary global bases, and a four-basis construction attains the minimum for arbitrary input states. For these nine observables, we also derive a variance lower bound valid for any number of global projective settings within the fixed-allocation, unbiased linear estimation model. A numerically optimized four-setting design has a Haar-averaged variance approximately $0.0341\%$ above this task-specific bound.

We show that shared measurement bases can improve precision for structured sets of observables. In Table~\ref{tab:comparison}, the comparison implementation with the lowest variance, which reuses data from all target eigenbases, requires about $4.1$ times as many copies for the same variance averaged over observables and Haar-random pure states. The six-qubit examples exhibit a finite-copy crossover. The shared designs have lower average RMSE over a range of sample budgets despite reconstruction bias; for stricter target errors, spectral reuse becomes more efficient as the shared error approaches its residual floor. These numerical comparisons use the specified measurement designs and estimator families.

%% file: problem_formulation.tex
A measurement setting applies a fixed unitary $U_i$ before an orthogonal projective measurement in the computational basis. We denote its measurable traceless operator space by $A(U_i)$. The same programmable measurement device may implement several settings. For $n$ qubits, the Hilbert-space dimension is $d=2^n$. Write $[n]=\{1,\ldots,n\}$. For a nonempty subset $S\subseteq[n]$, define $Z_S=\prod_{q\in S}Z_q$, where $Z_q$ acts as Pauli $Z$ on qubit $q$ and as the identity on the other qubits. We then have
\begin{equation}
\begin{aligned}
A(U_i)
&=\operatorname{span}_{\mathbb R}
\{U_i^\dagger Z_SU_i:\varnothing\ne S\subseteq[n]\}\\
&=U_i^\dagger\mathcal D_0U_i.
\end{aligned}
\label{eq:measurable-space}
\end{equation}
Here, $\operatorname{span}_{\mathbb R}$ denotes the real linear span, and $\mathcal D_0$ is the space of real diagonal $d\times d$ matrices with zero trace. Each measurement outcome is a bitstring from which we can compute the value of every $Z_S$. The same data therefore provide unbiased estimates of all observables in $A(U_i)$ for any input state.

Let $\mathcal O=\{O_1,\ldots,O_M\}$ be a set of $M$ Hermitian observables. We remove their identity components and define the traceless targets
\begin{equation}
\overline O_j=O_j-\frac{\operatorname{Tr}(O_j)}{d}I,
\qquad j=1,\ldots,M,
\label{eq:traceless-targets}
\end{equation}
where $I$ is the $d\times d$ identity matrix. The identity component contributes the known expectation value $\operatorname{Tr}(O_j)/d$, so we only need to measure $\overline O_j$. We write their real linear span as $V=\operatorname{span}_{\mathbb R}\{\overline O_1,\ldots,\overline O_M\}$.

If the spaces of $K$ settings satisfy
\begin{equation}
V\subseteq\sum_{i=1}^{K}A(U_i),
\label{eq:coverage}
\end{equation}
then we can write each traceless target as
\begin{equation}
\overline O_j=
\sum_{i=1}^{K}\sum_{\varnothing\ne S\subseteq[n]}
c_{ijS}\,U_i^\dagger Z_SU_i.
\label{eq:shared-decomposition}
\end{equation}
The real estimator coefficient $c_{ijS}$ gives the contribution of $U_i^\dagger Z_SU_i$ to $\overline O_j$. Setting $i$ measures the terms with index $i$, and we reuse its data for different targets $j$. These spaces may overlap, so the coefficients need not be unique.

Our goal is to minimize the number of settings. Allowing each $U_i$ to range over the $d\times d$ unitary matrices, we define
\begin{equation}
\begin{aligned}
K_\star(V)=\min\bigl\{K:\;&\exists\,U_1,\ldots,U_K,\\
&V\subseteq\sum_{i=1}^{K}A(U_i)\bigr\}.
\end{aligned}
\label{eq:minimum-devices}
\end{equation}
Here, $K$ is a nonnegative integer, with $K_\star(\{0\})=0$ when all targets contain only known identity components.

\subsection{Estimation and precision}

Besides the number of settings, we also consider the statistical error of the estimated expectation values. We first discuss exact reconstruction and then allow a residual error.

For a given decomposition, define the part of target $j$ measured in setting $i$ as
\begin{equation}
O_j^{(i)}=\sum_{\varnothing\ne S\subseteq[n]}
c_{ijS}U_i^\dagger Z_SU_i,
\qquad \overline O_j=\sum_{i=1}^{K}O_j^{(i)}.
\label{eq:device-components}
\end{equation}
We use $N$ independent copies of the same input state $\rho$, and assign a fixed number $N_i\geq1$ to each setting $i$ used in the experiment. Thus $N\geq K$, and the fraction $p_i=N_i/N$ satisfies $\sum_i p_i=1$. For a measurement outcome $x=(x_1,\ldots,x_n)$, where $x_q\in\{0,1\}$, write $\chi_S(x)=(-1)^{\sum_{q\in S}x_q}$. The contribution of this outcome to target $j$ is
\begin{equation}
f_{ij}(x)=\sum_{\varnothing\ne S\subseteq[n]}c_{ijS}\chi_S(x).
\label{eq:outcome-readout}
\end{equation}
Let $x_{it}$ be the outcome of the $t$th measurement in setting $i$. We estimate $\operatorname{Tr}(\rho O_j)$ by
\begin{equation}
\widehat o_j=\frac{\operatorname{Tr}(O_j)}{d}
+\sum_{i=1}^{K}\frac{1}{N_i}\sum_{t=1}^{N_i}f_{ij}(x_{it}).
\label{eq:unbiased-estimator}
\end{equation}
The decomposition ensures that $\mathbb E_\rho[\widehat o_j]=\operatorname{Tr}(\rho O_j)$ for any $\rho$.

Figure~\ref{fig:shared-measurements} summarizes the measurement procedure. We use the data from each setting for all observables, evaluate the estimator functions in Eq.~\eqref{eq:outcome-readout}, and combine their sample means using Eq.~\eqref{eq:unbiased-estimator}.
\input{measurement_figure}

Since different settings use independent copies, the variance is
\begin{equation}
\operatorname{Var}_\rho(\widehat o_j)
=\sum_{i=1}^{K}\frac{\operatorname{Var}_\rho(O_j^{(i)})}{N_i},
\label{eq:state-variance}
\end{equation}
where $\operatorname{Var}_\rho(Q)=\operatorname{Tr}(\rho Q^2)-[\operatorname{Tr}(\rho Q)]^2$ for a Hermitian operator $Q$. This expression includes the correlations between the estimator functions evaluated on the same bitstring.

To compare measurement schemes without knowing the input state, we average over pure states $\rho_\psi=|\psi\rangle\langle\psi|$ drawn uniformly from the Haar distribution, and give all $M$ targets equal weight. Define $C_i\in\mathbb R^{(d-1)\times M}$ by $(C_i)_{Sj}=c_{ijS}$, with one row for each nonempty $S$, and stack these matrices into $C$. We also write $U=(U_1,\ldots,U_K)$ and $p=(p_1,\ldots,p_K)$. The average variance takes the form
\begin{equation}
\begin{aligned}
\mathbb E_\psi\!\left[\frac{1}{M}\sum_{j=1}^{M}
\operatorname{Var}_{\rho_\psi}(\widehat o_j)\right]
&=\frac{\Gamma(U,C,p)}{N},\\
\Gamma(U,C,p)
&=\frac{d}{M(d+1)}\sum_{i=1}^{K}\frac{\|C_i\|_F^2}{p_i}.
\end{aligned}
\label{eq:variance-constant}
\end{equation}
Here, $\|C_i\|_F^2=\sum_{S,j}c_{ijS}^2$ is the squared Frobenius norm. The coefficient $\Gamma$ determines the average variance at a given number of state copies. We use the same measurement scheme for the entire state ensemble; the variance for an individual state can differ from this average.

For fixed $U$ and $C$, minimizing $\Gamma$ over the allocation gives
\begin{equation}
\begin{aligned}
p_i^\star&=\frac{\|C_i\|_F}{\sum_r\|C_r\|_F},\\
\Gamma^\star(U,C)
&=\frac{d}{M(d+1)}\left(\sum_{i=1}^{K}\|C_i\|_F\right)^2.
\end{aligned}
\label{eq:optimal-allocation}
\end{equation}
We omit settings with $C_i=0$. Thus, settings with larger estimator coefficients receive more copies. During optimization we treat $p_i$ as continuous; in an experiment we round the allocations to integers while keeping $\sum_i N_i=N$ and $N_i\geq1$ for each setting used. The variance formulas are exact for the actual fractions $p_i=N_i/N$. They assume ideal measurements. We keep the original normalization of every target observable.

\paragraph*{Approximate reconstruction.}
A nonzero reconstruction error can have two main sources. First, the targets may not admit an exact representation with the chosen small number of settings. Second, the search over $U_i$ may be restricted by the circuit depth or may stop at a local minimum or before convergence, even when an exact representation exists. A nonzero numerical loss alone does not distinguish these causes. We describe their effect through the reconstructed operators and residuals,
\begin{equation}
\begin{aligned}
\widetilde O_j
&=\frac{\operatorname{Tr}(O_j)}{d}I+\sum_{i=1}^{K}O_j^{(i)},\\
R_j&=O_j-\widetilde O_j,\qquad \operatorname{Tr}R_j=0.
\end{aligned}
\label{eq:approximate-residual}
\end{equation}
The estimator in Eq.~\eqref{eq:unbiased-estimator} now has mean $\operatorname{Tr}(\rho\widetilde O_j)$. Its bias and mean squared error (MSE), measured relative to the true expectation $\operatorname{Tr}(\rho O_j)$, are
\begin{equation}
\begin{aligned}
\operatorname{Bias}_\rho(\widehat o_j)
&=-\operatorname{Tr}(\rho R_j),\\
\operatorname{MSE}_\rho(\widehat o_j)
&=\operatorname{Var}_\rho(\widehat o_j)
  +[\operatorname{Tr}(\rho R_j)]^2.
\end{aligned}
\label{eq:bias-mse}
\end{equation}
The variance still follows Eq.~\eqref{eq:state-variance}, using the components actually measured in these settings. Averaging over targets and the same Haar pure-state ensemble gives
\begin{equation}
\begin{aligned}
\mathcal R_N(U,C,p)
&=\mathbb E_\psi\!\left[\frac1M\sum_{j=1}^{M}
  \operatorname{MSE}_{\rho_\psi}(\widehat o_j)\right]\\
&=\frac{\Gamma(U,C,p)}{N}
  +\frac{\sum_{j=1}^{M}\|R_j\|_F^2}{Md(d+1)}.
\end{aligned}
\label{eq:approximate-mse}
\end{equation}
Appendix~\ref{app:approximate-error} gives the derivation. At finite $N$, a smaller variance can compensate for a small reconstruction bias. For a fixed measurement scheme, the residual term remains as $N$ increases. Since it is independent of $p$ at fixed $U,C$, Eq.~\eqref{eq:optimal-allocation} still gives the optimal sample allocation. The minimum number of settings $K_\star(V)$ and the witnesses below continue to refer to exact reconstruction; we assess approximate reconstructions by their MSE.

\subsection{Classical variational optimization}

We first search for a decomposition with few settings, then minimize its variance. For each fixed $K$, we parameterize each $U_i$ as $U_i=\exp(iH_i)$, where $H_i$ is a Hermitian matrix whose independent real parameters we optimize. We can also use a variational circuit of fixed depth to parameterize $U_i$. Both choices preserve unitarity. We optimize all $K$ unitaries simultaneously.

To find a feasible decomposition, we minimize the reconstruction loss
\begin{equation}
\mathcal L_K(U)=\min_C
\frac{\displaystyle\sum_{j=1}^{M}
\left\|\overline O_j-\sum_{i=1}^{K}O_j^{(i)}\right\|_F^2}
{\displaystyle\sum_{j=1}^{M}\|\overline O_j\|_F^2}.
\label{eq:variational-loss}
\end{equation}
For operators, $\|Q\|_F^2=\operatorname{Tr}(Q^\dagger Q)$. We assume at least one nonzero traceless target; otherwise no setting is needed. At fixed $U$, the coefficients enter linearly, so we solve the inner least-squares problem by singular value decomposition. We recompute these coefficients as we optimize the unitaries. Eliminating the linear coefficients is a variable projection step \cite{Golub1973VariableProjection}. A zero loss means that the settings cover every target. In numerical calculations, we check both this loss and the reconstruction error of each target.

Starting from $K=1$, we optimize the unitaries from several initial conditions. If the loss remains above the chosen tolerance after the allowed number of iterations, we add a setting. We initialize the enlarged problem with the previous unitaries and one new unitary, then optimize all $K+1$ unitaries simultaneously. The search is nonconvex: failure to find a decomposition at a given $K$ does not prove that none exists. An exact decomposition gives an upper bound on $K_\star(V)$; the witnesses discussed later provide lower bounds.

Once we find a decomposition, we minimize the variance at fixed $K$ by solving
\begin{equation}
\begin{aligned}
\min_{U,C}\quad &\Gamma^\star(U,C)\\
\text{subject to}\quad
&\overline O_j=\sum_{i=1}^{K}O_j^{(i)},
\quad j=1,\ldots,M.
\end{aligned}
\label{eq:precision-design}
\end{equation}
Each $U_i$ remains unitary through its parameterization, and Eq.~\eqref{eq:optimal-allocation} gives the optimal $p$. At fixed $U$, minimizing $\sum_i\|C_i\|_F$ under these linear reconstruction constraints is a convex problem. When the decomposition is not unique, this step can improve upon the minimum-norm least-squares coefficients. We can also refine all $U_i$ together with $C$ under the reconstruction constraints. Fewer settings alone do not guarantee a smaller $\Gamma$; the estimator coefficients and sample allocation also matter.

For approximate reconstruction, we can instead choose the number of settings $K$, a circuit ansatz, and the number of state copies $N$, and minimize
\begin{equation}
\begin{aligned}
\min_{\theta,C,p}\quad &\mathcal R_N(U(\theta),C,p)\\
\text{subject to}\quad &p_i>0,\qquad \sum_i p_i=1.
\end{aligned}
\label{eq:mse-design}
\end{equation}
Here, $\theta$ contains the circuit parameters of all settings. We omit settings with zero coefficient blocks as above. At fixed $U,p$, the problem in $C$ is a convex quadratic problem, with the closed-form solution in Eqs.~\eqref{eq:finite-copy-readout} and~\eqref{eq:finite-copy-risk}. At fixed $U,C$, the allocation again follows Eq.~\eqref{eq:optimal-allocation}. We optimize all measurement unitaries simultaneously while updating the estimator coefficients and sample allocation. Minimizing this objective may introduce a small bias to reduce the variance at finite $N$; the nonconvex search over $\theta$ need not reach its global minimum. Appendix~\ref{app:initialization-checks} reports additional initialization checks for the fixed examples; these do not establish a general success probability.

We perform this optimization classically before the experiment, using the target operators and the chosen error criterion. After obtaining $U$, $C$, and $p$, we implement each setting through its fixed unitary and computational-basis readout, allocate the state copies, and estimate the expectation values from the outcomes using Eq.~\eqref{eq:unbiased-estimator}.

Quantum channels provide another variational approach to observable estimation \cite{Yang2025Incompatible}.

%% file: measurement_figure.tex
\begin{figure}[t]
\centering
\includegraphics[width=\columnwidth]{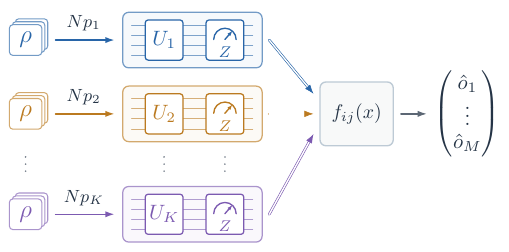}
\caption{Shared estimation with $K$ fixed measurement settings. We assign $N_i=Np_i$ independent copies of the same state $\rho$ to setting $i$. In setting $i$, we apply $U_i$ and then perform an orthogonal projective measurement in the computational basis, indicated by the meter labeled $Z$. The double lines carry classical measurement outcomes. For every target $O_j$, we use the coefficients $c_{ijS}$ collected in $C$ to evaluate the estimator function $f_{ij}(x)$ in Eq.~\eqref{eq:outcome-readout}. We average these values separately for each setting, sum the contributions from all settings, and add the known identity component $\operatorname{Tr}(O_j)/d$, as in Eq.~\eqref{eq:unbiased-estimator}. We use the same measurement data for all observables. We show the first two and last settings; the four wires only illustrate a multiqubit input.}
\label{fig:shared-measurements}
\end{figure}

%% file: witnesses.tex
\subsection{Lower bounds from witnesses}
\label{sec:witnesses}

A successful variational search gives an upper bound on the number of measurement settings needed. To determine whether fewer settings could suffice, we also need lower bounds. A first bound follows by counting dimensions. Let $m=\dim_{\mathbb R}V$ be the dimension of the traceless target space. Since each setting contributes a space of dimension $d-1$, any feasible $K$ satisfies
\begin{equation}
m\le\dim\!\left(\sum_{i=1}^K A(U_i)\right)\le K(d-1).
\end{equation}
Therefore,
\begin{equation}
K_\star(V)\ge\left\lceil\frac{m}{d-1}\right\rceil.
\label{eq:dimension-bound}
\end{equation}
This bound depends on the dimension $m$ of the target space, which can be smaller than the number $M$ of observables.

The dimension bound does not use the structure of the observables and can therefore be loose. To describe such structure, choose a real linear space
\begin{equation}
\begin{aligned}
V&\subseteq\mathcal W\subseteq\operatorname{Herm}_0(d),\\
\operatorname{Herm}_0(d)&=\{X=X^\dagger:\operatorname{Tr}X=0\}.
\end{aligned}
\label{eq:structural-space}
\end{equation}
For example, $\mathcal W$ may describe locality constraints or common symmetries. We only require the measurement spaces to cover $V$, not all of $\mathcal W$. Neither space needs to be closed under products or commutators. Define the internal center
\begin{equation}
\begin{aligned}
F=Z(\mathcal W)=\{Y\in\mathcal W:\;&[Y,X]=0\\
&\text{for all }X\in\mathcal W\},
\end{aligned}
\label{eq:internal-center}
\end{equation}
and the parameters
\begin{equation}
\begin{aligned}
\alpha(\mathcal W)
&=\max_{\substack{L\subseteq\mathcal W\\ [L,L]=0}}
\dim_{\mathbb R}L,\\
c&=\dim_{\mathbb R}F,\qquad
h=\dim_{\mathbb R}(V\cap F).
\end{aligned}
\label{eq:structural-parameters}
\end{equation}
Here, $L$ is a real linear subspace, and $[L,L]=0$ means pairwise commutation. Thus $\alpha(\mathcal W)$ counts the largest number of independent commuting directions in $\mathcal W$, including arbitrary linear combinations. The numbers $c$ and $h$ count the central directions in $\mathcal W$ and those already contained in $V$, respectively. They obey $0\le h\le c\le\alpha(\mathcal W)\le d-1$. The center is internal: an operator outside $\mathcal W$ is not counted in $F$, even if it commutes with all of $\mathcal W$.

\Needspace{4\baselineskip}
\begin{theorem}[Structural lower bound]
\label{thm:structural-capacity}
If $K\ge1$ measurement settings cover $V\subseteq\mathcal W$, then
\begin{equation}
m\le(K-1)(d-1)+\alpha(\mathcal W)-c+
             \left\lfloor\frac hK\right\rfloor.
\label{eq:structural-bound}
\end{equation}
If a positive integer $K$ violates this inequality, then $K_\star(V)\ge K+1$.
\end{theorem}

Appendix~\ref{app:proof-structural} gives the proof. The measurement bases remain arbitrary and global, and the input state is unrestricted. In particular, the measurement spaces may contain directions outside $\mathcal W$, and their sum may strictly contain $V$. The inequality is necessary; satisfying it does not establish a construction with $K$ settings. A certified upper bound on $\alpha(\mathcal W)$ can replace its exact value and gives a possibly weaker test. Different choices of $\mathcal W$ may yield different bounds; we can choose the one that gives the strongest bound.

Common symmetries provide one useful choice. If Hermitian operators $Q_1,\ldots,Q_t$ commute with every target, choose
\begin{equation}
\mathcal W=\{X\in\operatorname{Herm}_0(d):
[Q_\mu,X]=0\ \text{for all }\mu\}.
\label{eq:common-commutant}
\end{equation}
The $Q_\mu$ need not commute with one another. Theorem~\ref{thm:structural-capacity} then gives the common-symmetry lower bound; Appendix~\ref{app:proof-structural} explains how to calculate its parameters from the block structure. Choosing $\mathcal W=\operatorname{Herm}_0(d)$ recovers Eq.~\eqref{eq:dimension-bound}. This theorem alone can improve the dimension lower bound by at most one setting. Appendix~\ref{app:four-device-example} gives a two-qubit example in which the structural bound and an explicit construction establish that exactly four settings are needed.

We next construct a witness that distinguishes two settings from three. We evaluate it from the target operators without optimizing the measurement bases. Define
\begin{equation}
\begin{aligned}
\tau_q(X_1,\ldots,X_q)
={}&\frac1{q!}\sum_{\pi\in S_q}\operatorname{sgn}(\pi)\\[-2pt]
&\times\operatorname{Tr}\!\left(X_{\pi(1)}\cdots X_{\pi(q)}\right).
\end{aligned}
\label{eq:alternating-trace}
\end{equation}
Here, $S_q$ is the set of permutations of $q$ labels, and $\operatorname{sgn}(\pi)$ is the sign of the permutation.

\begin{theorem}[Witness for at least three settings]
\label{thm:odd-trace-witness}
If $K_\star(V)\le2$, then, for every odd integer $q\ge3$,
\begin{equation}
\tau_q(X_1,\ldots,X_q)=0
\qquad\text{for all }X_1,\ldots,X_q\in V.
\label{eq:two-device-vanishing}
\end{equation}
Any nonzero value therefore proves \mbox{$K_\star(V)\ge3$}.
\end{theorem}

The proof uses only commutation within each measured space and cyclicity of the trace; it does not require the target space to be closed under commutators. We give the proof in Appendix~\ref{app:proof-odd-trace}.

The simplest test uses three operators. For Hermitian $X,Y,Z$, define the real number
\begin{equation}
\Omega_0(X,Y,Z)
=-\frac{i}{d}\operatorname{Tr}\!\left(X[Y,Z]\right)
=-\frac{2i}{d}\tau_3(X,Y,Z).
\label{eq:cubic-witness}
\end{equation}
A nonzero $\Omega_0$ rules out every representation using two settings, including decompositions in which each target uses both settings. If all cubic values vanish, higher odd orders can still detect the need for more than two settings. Vanishing witnesses alone do not establish that a representation using two settings exists.

The structural bound applies to any number of measurement settings and can prove that some sets of observables require four or more settings. Obtaining tighter bounds for general target spaces, or extending the alternating-trace construction to equally simple witnesses that exclude three or more settings, remains an open question.

These bounds complement the variational construction: a witness gives a lower bound, while a feasible decomposition gives an upper bound. When both bounds agree, they determine the minimum number of measurement settings. The next section illustrates these bounds with collective moments and multiqubit examples.

%% file: collective_example.tex
We first consider the collective first and second moments of two qubits.
Define
\begin{equation}
\begin{aligned}
J_\alpha&=\tfrac12(\sigma_\alpha\otimes I_2+I_2\otimes\sigma_\alpha),\\
Q_{\alpha\alpha}&=J_\alpha^2-\tfrac12I_4
=\tfrac12\sigma_\alpha\otimes\sigma_\alpha,\\
Q_{\alpha\beta}&=\tfrac12\{J_\alpha,J_\beta\}\\
&=\tfrac14(\sigma_\alpha\otimes\sigma_\beta+
\sigma_\beta\otimes\sigma_\alpha),\quad\alpha<\beta,
\end{aligned}
\label{eq:collective-targets}
\end{equation}
where $\alpha,\beta\in\{x,y,z\}$. The three first moments and
six second moments are traceless and linearly independent.
They commute with swap and span
\begin{equation}
\begin{aligned}
V=\mathcal W&=\operatorname{Herm}_0(M_3(\mathbb C)\oplus\mathbb C),\\
\dim_{\mathbb R}V&=9,
\end{aligned}
\label{eq:collective-target-space}
\end{equation}
where the two blocks act on the triplet and singlet sectors.

Dimension counting gives only $K\ge3$. The block structure gives
$\alpha(\mathcal W)=3$ and $c=h=1$, so three settings would require
\begin{equation}
9\le(3-1)\cdot3+3-1+\lfloor1/3\rfloor=8
\end{equation}
by Theorem~\ref{thm:structural-capacity}. This excludes every choice
of three global bases, including bases that mix the two sectors.
For a matching construction, take four mutually unbiased bases (MUBs) in
the three-dimensional triplet sector and add the singlet to each
\cite{Wootters1989MUB}. Their triplet-traceless parts form four
orthogonal two-dimensional spaces. The full measurement spaces also
contain a common central direction, so their sum covers all nine
target directions.
Hence
\begin{equation}
K_\star(V)=4.
\label{eq:collective-exact-count}
\end{equation}
Appendix~\ref{app:four-device-example} gives the explicit bases
and proof.

Standard permutationally invariant (PI) tomography based on common
local spin axes uses six settings for two qubits
\cite{Toth2010,Moroder2012}.
Klimov \textit{et al.}\ reduce a MUB measurement collection to four
bases using a prior of permutation invariance of the input state
\cite{Klimov2013}.
Our four-setting result estimates these nine observables for
arbitrary input states on the full four-dimensional space.

We assess precision separately, keeping the natural normalization
of Eq.~\eqref{eq:collective-targets}. For equal target weights and
the Haar pure-state average on the full four-dimensional space,
the average variance is $\Gamma/N$. Every exact unbiased linear
reconstruction with fixed allocation obeys
\begin{equation}
\Gamma\ge\frac{31}{60}+\frac{\sqrt2}{5}
=0.7995093791\ldots .
\label{eq:collective-main-variance-bound}
\end{equation}
This bound allows any finite number of global projective settings,
including bases that mix the two sectors; Appendix~\ref{app:collective-global-bound}
gives the proof.

The explicit triplet-MUB construction gives $\Gamma=0.847364$
after readout and allocation optimization. Varying all four triplet
bases while keeping the singlet fixed, and optimizing the readout
and allocation, gives the numerical value
\begin{equation}
\Gamma_{4,\mathrm{num}}\simeq0.7997819463.
\end{equation}
For these nine targets, this value is approximately $0.0341\%$
above Eq.~\eqref{eq:collective-main-variance-bound}.
Table~\ref{tab:collective-comparison} gives the comparison protocols.

%% file: example.tex
We next consider a set of $M=30$ observables on four qubits ($d=16$). We use this example to show how shared measurement bases can reduce both the number of measurement settings and the number of state copies. We construct the observables from three independently drawn Haar-random unitaries $G_1,G_2,G_3$:
\begin{equation}
\begin{aligned}
\widetilde O_j
&=\sum_{r=1}^{3}\sum_{\varnothing\ne S\subseteq[4]}
w_{rSj}\,G_r^\dagger Z_SG_r,\\
O_j&=\frac{\widetilde O_j}{\sqrt{3}\,\|\widetilde O_j\|_0},
\qquad \|Q\|_0^2=\frac{\operatorname{Tr}(Q^\dagger Q)}{16}.
\end{aligned}
\label{eq:example-targets}
\end{equation}
We use a dense real $45\times30$ coefficient matrix $W=(w_{rSj})$ with full column rank and different scales for its three blocks. Thus the targets mix contributions from three independently chosen bases. All targets are traceless, and $\operatorname{Tr}(O_j^2)/16=1/3$. Their span has dimension $30$. We fix the random seed to $7501$ before comparing the measurement schemes. Appendix~\ref{app:shared} specifies the construction, and the source archive contains the target matrices.

\begin{table}[!tbp]
\caption{Average variance constants for the same $30$ observables. The mean variance over observables and Haar-random pure input states is $\Gamma/N$. The last column gives the number of copies relative to our scheme at the same mean variance. All schemes except global shadows use fixed sample allocation; global shadows draws measurements independently. We report values for continuous allocation fractions, before rounding the number of copies to integers.}
\label{tab:comparison}
\begin{ruledtabular}
\begin{tabular}{lrr}
Scheme & $\Gamma$ & Relative copies\\
\hline
Shared three bases & $0.828049$ & $1.00$\\
Independent spectral & $9.411765$ & $11.37$\\
Spectral reuse & $3.393722$ & $4.10$\\
Pauli GC & $5.239637$ & $6.33$\\
Rotated joint measurement & $4.796128$ & $5.79$\\
Global Haar/Clifford shadows & $5.647059$ & $6.82$\\
\end{tabular}
\end{ruledtabular}
\end{table}

Dimension counting gives only $K_\star(V)\ge\lceil30/15\rceil=2$. The cubic witness is stronger: for three targets, we obtain
\begin{equation}
\Omega_0(O_{15},O_{16},O_{28})\simeq-0.02508950\ne0.
\label{eq:example-witness}
\end{equation}
Theorem~\ref{thm:odd-trace-witness} therefore gives $K_\star(V)\ge3$. Equation~\eqref{eq:example-targets} provides a construction with three settings, so
\begin{equation}
K_\star(V)=3.
\label{eq:example-minimum}
\end{equation}
Appendix~\ref{app:example-certificate} specifies the exact construction behind the finite-precision data and certifies that the witness stays nonzero after accounting for rounding.

For the same exact instance, restricting every measurement basis to a
Clifford basis in the original Pauli coordinates requires $17$ settings
(Appendix~\ref{app:clifford-count}).

We optimize using only the target observables, without the unitaries or coefficients used to construct them. For each number of measurement settings, we jointly optimize all measurement unitaries and solve for the linear coefficients. We obtain a decomposition over three settings with maximum relative reconstruction error
\begin{equation}
\max_j\frac{\|O_j-\sum_{i=1}^3O_j^{(i)}\|_F}{\|O_j\|_F}
=3.48\times10^{-14}.
\label{eq:example-residual}
\end{equation}
For this solution, the measurement bases uniquely determine the estimator coefficients. After relabeling the settings, Eq.~\eqref{eq:optimal-allocation} gives
\begin{equation}
(p_1,p_2,p_3)=(0.189751,\,0.309419,\,0.500830),
\label{eq:example-allocation}
\end{equation}
and the resulting variance constant is $\Gamma_{\rm shared}=0.828049$. Our variational search thus recovers the minimum number of measurement settings established by the witness and the construction. We report the variance of this optimized measurement scheme, without claiming a global minimum over all choices of the three bases.

We compare with independent spectral measurements \cite{Watrous2018}, spectral measurements with data reuse, globally commuting Pauli groups (Pauli GC) \cite{Yen2020Compatible}, a globally rotated local joint measurement \cite{McNulty2023Joint}, and global classical shadows \cite{Huang2020Shadows}. For spectral reuse, we use every target eigenbasis to estimate all targets and optimize the readout and allocation \cite{DAriano2007Optimal,Fischer2024DualFrame}. For Pauli GC, we optimize the allocation over all $2295$ maximal commuting groups, reusing every occurrence of each term. For the joint protocol, we optimize the global pre-rotation, local strengths, readout, and allocation over $256$ product settings. The joint-measurement search is nonconvex. For global shadows, we average the canonical snapshots. Uniform global Haar and uniform global Clifford measurements have the same variance in this comparison because the multiqubit Clifford group is a unitary $3$-design \cite{Zhu2017Designs}. Appendix~\ref{app:protocols} describes the measurements and estimators.

Table~\ref{tab:comparison} shows that spectral reuse is the strongest comparison scheme here. Its optimized variance constant is $3.393722$, compared with $0.828049$ for our three shared bases. It therefore requires about $4.10$ times as many copies for the same average variance. At $N=10^4$, for example, our mean variance is $8.28\times10^{-5}$, compared with $3.39\times10^{-4}$ for spectral reuse. These ratios refer to the stated average over input states and need not hold for each individual state.

This example illustrates the benefit of choosing measurement bases that exploit the structure of the observables. The gain depends on the decomposition coefficients and sample allocation, as well as the number of measurement settings. We compare ideal measurements and total numbers of copies here. Our comparison does not include the circuit cost of implementing arbitrary global transformations.

%% file: six_qubit_figure.tex
\begin{figure*}[t]
\centering
\begin{minipage}[t]{0.49\textwidth}
\raggedright\textbf{(a)}\par\vspace{-0.9\baselineskip}
\centering
\includegraphics[width=\linewidth]{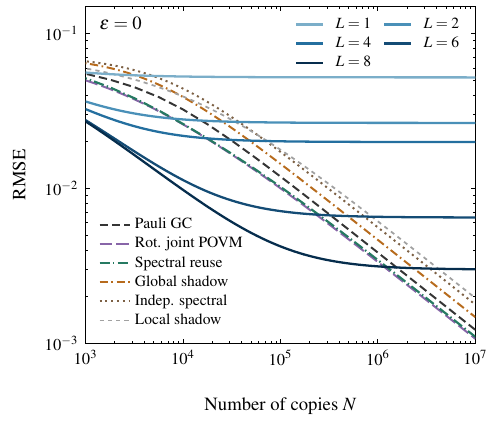}
\end{minipage}\hfill
\begin{minipage}[t]{0.49\textwidth}
\raggedright\textbf{(b)}\par\vspace{-0.9\baselineskip}
\centering
\includegraphics[width=\linewidth]{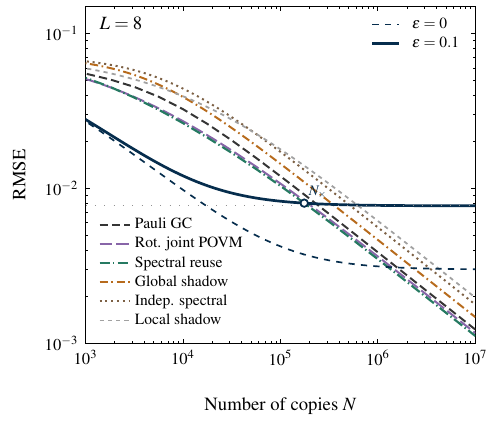}
\end{minipage}
\caption{Finite-depth measurements of $96$ six-qubit observables. RMSE uses Eq.~\eqref{eq:approximate-mse}. (a) Clean targets from three structured $20$-block circuits, measured in three settings with $L=1,2,4,6,8$ blocks. (b) Perturbed targets at $\epsilon=0.1$, $L=8$; the dashed blue reference uses the optimized clean instance. For shared designs, we fix the unitaries trained at $N_0=10^4$ and optimize $C,p$ at each evaluated $N$. Spectral reuse and Pauli GC use separately converged allocations at their evaluated budgets; lines connect these nodes as guides to the eye. Appendix~\ref{app:baseline-convergence} reports the convergence checks. The dotted line and circle mark the perturbed residual floor and the observed crossover. The comparison uses the stated readout families and continuous quotas.}
\label{fig:six-qubit-results}
\end{figure*}

%% file: six_qubit_example.tex
We next examine whether our scheme retains an advantage when shallow measurement circuits leave a small reconstruction error. We construct $M=96$ traceless observables on six qubits ($d=64$) from three independent circuits $G_1,G_2,G_3$:
\begin{equation}
\begin{aligned}
H_j&=\sum_{r=1}^{3}\sum_{\varnothing\ne S\subseteq[6]}
w_{rSj}G_r^\dagger Z_SG_r,\\
O_j^{(0)}&=\frac{H_j}{\sqrt{3}\,\|H_j\|_0},
\qquad \|Q\|_0^2=\frac{\operatorname{Tr}(Q^\dagger Q)}{64}.
\end{aligned}
\label{eq:six-targets}
\end{equation}
The mixing coefficients form a dense real $189\times96$ matrix. Each $G_r$ contains $20$ rotation blocks, each with a parallel $R_z$ layer followed by a parallel $R_y$ layer. Alternating nearest-neighbor CZ matchings on a ring connect successive blocks. The first two blocks use independent angles uniform on $[-\pi,\pi]$; the remaining eighteen use independent Gaussian angles of standard deviation $0.08$. We thus use structured circuits with large rotations in the first two blocks and weak rotations thereafter, rather than generic deep random circuits. The $20$ blocks specify our construction; we do not establish the minimum circuit depth. Likewise, the exact construction with three settings gives an upper bound on the number of settings; we do not certify its minimum in this example.

We fix $K=3$ and jointly optimize the measurement circuits using $L=1,2,4,6,8$ rotation blocks of the same architecture. Each measurement circuit has $12L$ rotation parameters and $3(L-1)$ CZ gates. The classical optimization uses only the target observables. We minimize the mean squared error in Eq.~\eqref{eq:mse-design} at $N_0=10^4$, then keep the optimized $U_i$ fixed and reoptimize $C,p$ for each plotted $N$. Changing the number of copies therefore does not require a new measurement circuit. Appendix~\ref{app:six-qubit} gives the construction and comparison details.

Figure~\ref{fig:six-qubit-results}(a) shows the absolute root mean squared error, $\mathrm{RMSE}=\sqrt{\mathcal R_N}$, averaged over observables and Haar-random pure input states. We compare with independent spectral measurements, spectral measurements with full data reuse, Pauli GC, the globally rotated local joint measurement, and global classical shadows. We also include uniform local Pauli shadows~\cite{Huang2020Shadows}. For this finite-sample comparison, we optimize the linear readout within each stated estimator family, rather than applying only one shrinkage factor to each target~\cite{Cai2025Biased,Fischer2024DualFrame}. In particular, spectral reuse combines all $96$ target eigenbases through a full linear readout, while the Pauli-based protocols allow a separate shrinkage factor for each Pauli coefficient. We optimize the spectral-reuse and Pauli GC allocations at each evaluated budget, using stopping criteria selected from convergence checks (Appendix~\ref{app:baseline-convergence}). The other comparison implementations retain the measurement parameters and allocation families specified in Appendix~\ref{app:six-qubit}. At $N=10^4$, our scheme with $L=8$ reaches an RMSE of $9.79\times10^{-3}$, compared with $2.59\times10^{-2}$ for the comparison implementation with the lowest RMSE at this budget, the rotated joint measurement. Our RMSE is smaller by a factor of $2.65$. Among the tested designs, larger circuit depths lower the asymptotic error and extend the range of precisions over which our scheme has an advantage. A finite residual remains: the optimized $L=8$ scheme approaches an RMSE of $2.99\times10^{-3}$ as $N\to\infty$.

To test imperfect shared structure, we also use
\begin{equation}
O_j^{(\epsilon)}=
\frac{O_j^{(0)}+\epsilon\Delta_j}{\sqrt{1+\epsilon^2}},
\label{eq:six-perturbed-targets}
\end{equation}
where each $\Delta_j$ is a random traceless Hermitian operator orthogonal to the original generating space, with $\operatorname{Tr}(\Delta_j^2)/64=1/3$. This normalization keeps $\operatorname{Tr}[(O_j^{(\epsilon)})^2]/64=1/3$. We separately optimize the measurement scheme for the perturbed observables at $\epsilon=0.1$, using $L=8$. Here $\epsilon$ measures the relative perturbation of an operator, rather than an absolute error in its expectation value.

Figure~\ref{fig:six-qubit-results}(b) shows that our scheme retains an advantage despite a small bias. At $N=10^4$, the RMSE is $1.21\times10^{-2}$, compared with $2.64\times10^{-2}$ for spectral measurements with full data reuse, the comparison implementation with the lowest RMSE at this budget. In the displayed range, the calculated curves show a lower RMSE for our scheme before the observed crossover, where
\begin{equation}
N_\times\simeq 1.75\times10^5,
\qquad \mathrm{RMSE}(N_\times)\simeq 8.04\times10^{-3}.
\label{eq:six-crossing}
\end{equation}
This crossing concerns the fixed shared design and the specified comparison implementations in Fig.~\ref{fig:six-qubit-results}. Independent solves near the crossing retain the ordering on opposite sides after accounting for the numerical allocation gaps (Appendix~\ref{app:baseline-convergence}). In the displayed curves, target RMSE values above the crossing value require fewer copies with the shared design; for stricter error targets, spectral reuse becomes more efficient. The shared design approaches a residual floor of $7.72\times10^{-3}$. This residual reflects restricted depth and incomplete convergence; it is not a lower bound for all circuits of that depth. Comparisons use ideal gates and readout, the stated state average, and the specified readout families.

%% file: conclusion.tex
We study the minimum number of orthogonal projective measurement settings needed to estimate multiple expectation values. We combine a classical variational search with structural and alternating-trace witnesses. Matching constructions certify exact minimum setting counts beyond dimension counting. For our four-qubit example, exactly three settings suffice, and the optimized shared measurements reduce the Haar-averaged variance relative to the comparison protocols.

For the collective first and second moments of two qubits, the structural bound and a four-basis construction certify the minimum setting count for arbitrary input states. For this task, a numerically optimized four-basis design has a Haar-averaged variance within approximately $0.0341\%$ of a bound valid for any number of global projective settings, under the fixed-allocation, unbiased linear estimation model used here.

For the perturbed six-qubit instance, the plotted shared design has a lower target- and Haar-averaged RMSE than the specified comparison implementations before the observed crossover. For target errors below the crossover value, spectral reuse becomes more efficient as the shared design approaches its residual floor. This comparison concerns the fixed measurement designs and estimator families studied here.

Several questions remain open. Tighter structural bounds and extensions of the alternating-trace witness that rule out three or more settings would help determine the minimum number for more general sets of observables. More efficient classical optimization could make the method applicable to larger systems. Including gate and readout noise, with suitable calibration \cite{Koh2022NoisyShadows,Chen2021RobustShadows}, would also help assess its performance under experimental conditions.

%% file: acknowledgements.tex
This work was supported by the Science Challenge Project (Grant No.~TZ2025017), and the National Key Research and Development Program of China (Grant Nos.~2021YFA0718302 and 2021YFA1402104).

%% file: measurement_protocols.tex
We first specify the protocols used in Table~\ref{tab:comparison}, with $n=4$, $d=16$, and $M=30$. Appendix~\ref{app:six-qubit} describes the same protocols for the six-qubit example. Let $\mathcal P_n=\{I,X,Y,Z\}^{\otimes n}$ denote the Hermitian Pauli strings, with $\operatorname{Tr}(PP')=d\,\delta_{PP'}$. Write
\begin{equation}
\begin{aligned}
O_j&=\sum_{P\ne I}a_{jP}P,\\
a_{jP}&=\frac{\operatorname{Tr}(PO_j)}{d},\qquad
e_P=\sum_j a_{jP}^2.
\end{aligned}
\label{eq:pauli-coefficients}
\end{equation}
When the decompositions are exact, all protocols give unbiased estimates of every target for any input state. We calculate their variances using the same average over Haar-random pure states as in Eq.~\eqref{eq:variance-constant}. For fixed sample allocations, we treat the numbers of copies as continuous; an experiment requires rounding them to integers.

\subsection{Shared measurement bases and target construction}
\label{app:shared}

In our shared-basis scheme, for setting $i$ we apply the optimized $U_i$ and measure in the computational basis. From the resulting bitstring, we compute all $15$ nontrivial $Z$ parities, combine them with $c_{ijS}$, and use Eq.~\eqref{eq:unbiased-estimator}. We allocate copies according to Eq.~\eqref{eq:optimal-allocation}. This uses three fixed orthogonal projective measurements. The decomposition into rotated diagonal spaces is related to the commuting-fragment approach of Ref.~\cite{Yen2021Cartan}; here we use the same optimized settings for all targets. Our earlier work uses a unital quantum channel as a multiple-output quantum neural network to estimate incompatible observables \cite{Yang2025Incompatible}. Here we choose measurement bases on the input system and combine their data classically.

For completeness, we generate the real mixing matrix in Eq.~\eqref{eq:example-targets} as
\begin{equation}
\begin{aligned}
W&=\Lambda L\,\operatorname{diag}(t_1,\ldots,t_{30})R^{\mathsf T},\\
\Lambda&=\operatorname{diag}(I_{15},1.7I_{15},2.8I_{15}).
\end{aligned}
\label{eq:mixing-matrix}
\end{equation}
Here, $L$ is the $45\times30$ orthonormal factor of a real Gaussian matrix, $R$ is an independently generated $30\times30$ orthogonal factor, and the $t_a$ are geometrically spaced from $0.7$ to $1.3$. We draw the three complex Haar unitaries independently and normalize each target as in Eq.~\eqref{eq:example-targets}. The fixed target matrices, optimized measurement unitaries, and estimator coefficients are included with the manuscript sources.

To find the measurement bases, we fit an orthonormal basis for the span of the target observables and solve the linear least-squares problem at each joint update of the unitaries. This choice changes how we weight a nonzero reconstruction error but preserves the condition for exact reconstruction in Eq.~\eqref{eq:coverage}. We then recover the coefficients for the original, normalized targets to evaluate $\Gamma$. The $255\times45$ Pauli-coordinate measurement matrix of the solution used here has full column rank, with smallest singular value $0.656726$. Its estimator coefficients are therefore unique at fixed $U$. Their block norms, in the order of Eq.~\eqref{eq:example-allocation}, are $(0.974848,1.589643,2.573023)$. This numerical result does not establish a global minimum over all unitaries.

\subsection{Independent spectral measurements}
\label{app:spectral}

We diagonalize each target,
\begin{equation}
O_j=\sum_b\lambda_{jb}|v_{jb}\rangle\langle v_{jb}|,
\end{equation}
and implement the corresponding orthogonal projective measurement \cite{Watrous2018}. We assign $N_j=Nq_j$ copies to target $j$ and average its observed eigenvalues. In this protocol, we estimate each target from separate measurement data. The Haar-averaged single-copy variance of a traceless target is $h_j=\operatorname{Tr}(O_j^2)/(d+1)$. Thus
\begin{equation}
\begin{aligned}
\Gamma_{\rm spec}&=\frac1M\sum_j\frac{h_j}{q_j},\\
q_j^\star&=\frac{\sqrt{h_j}}{\sum_k\sqrt{h_k}},\\
\Gamma_{\rm spec}^\star&=\frac1M\left(\sum_j\sqrt{h_j}\right)^2
=\frac{160}{17}.
\end{aligned}
\label{eq:spectral-risk}
\end{equation}
All targets have the same norm in this example, so $q_j^\star=1/30$.

\subsection{Spectral measurements with data reuse}
\label{app:spectral-reuse}

Independent spectral measurement discards the information that one target's eigenbasis may provide about the others. We therefore use a stronger protocol with the same $M$ target eigenbases. Let $V_r$ have the eigenvectors of $O_r$ as columns. For setting $r$, we apply $U_r=V_r^\dagger$ and measure in the computational basis. We retain its full outcome $b$ and use it for every target, assigning a separate score $f_{rj}(b)$ for each $j$. With $N_r=Np_r$ copies, the estimator is
\begin{equation}
\widehat o_j=\sum_{r=1}^{M}\frac1{N_r}
   \sum_{t=1}^{N_r}f_{rj}(b_{rt}).
\label{eq:spectral-reuse-estimator}
\end{equation}
We choose centered scores, $\sum_b f_{rj}(b)=0$, and require
\begin{equation}
O_j=\sum_{r,b} f_{rj}(b)
       |v_{rb}\rangle\langle v_{rb}|.
\label{eq:spectral-reuse-decomposition}
\end{equation}
This identity makes every estimate unbiased for any input state. Branch-dependent constant scores do not change the variance; for traceless targets we can set them to zero without losing an estimator. Expanding each centered score in the $d-1$ nontrivial $Z$ parities gives the coefficient blocks $C_r$ of Eq.~\eqref{eq:variance-constant}. We optimize these blocks and the allocation while keeping the eigenbases fixed, using the freedom of linear measurement reconstruction \cite{Scott2006Tight,DAriano2007Optimal,Fischer2024DualFrame}.

For clarity, this fixed-basis optimization has a direct weighted least-squares solution. Choose orthonormal real coordinates for $\operatorname{Herm}_0(d)$ with inner product $\operatorname{Tr}(AB)/d$. Let $\mathsf A_r$ contain the coordinate vectors of $U_r^\dagger Z_SU_r$ as columns, and let $T$ contain the target vectors. Set
\begin{equation}
F(p)=\sum_r p_r\mathsf A_r\mathsf A_r^{\mathsf T}.
\label{eq:fixed-basis-frame}
\end{equation}
For positive allocations and targets in the measurement span, the optimal coefficients and variance are
\begin{equation}
\begin{aligned}
C_r^\star&=p_r\mathsf A_r^{\mathsf T}F(p)^+T,\\
\Gamma_{\rm lin}(p)
&=\frac{d}{M(d+1)}
  \operatorname{Tr}\!\left[T^{\mathsf T}F(p)^+T\right].
\end{aligned}
\label{eq:fixed-basis-optimal-readout}
\end{equation}
Here $F(p)^+$ is the Moore--Penrose inverse on the measurement span. We minimize this convex function over the allocation simplex. Equivalently, the joint problem in $C,p$ minimizes the quadratic perspectives $\sum_r\|C_r\|_F^2/p_r$ under the linear reconstruction constraints. This procedure also applies to any other fixed pool of orthogonal projective measurements.

For the four-qubit targets, the $30$ eigenbases span all $255$ traceless directions. The optimized coefficients and allocation give $\Gamma_{\rm reuse}=3.393722$, compared with $3.404788$ at equal allocation and $9.411765$ when each target uses only its own spectral data. Thus most of this improvement comes from reusing the outcomes. The optimization fixes the target eigenbases; it does not search over arbitrary new measurement bases.

\subsection{Globally commuting Pauli groups}
\label{app:gc}

We use the standard fact that a Clifford transformation can simultaneously diagonalize a commuting group of Pauli strings \cite{Yen2020Compatible,Gokhale2020MeasurementCost,Reggio2024Partitioning}. Our candidate pool consists of all $2295$ maximal commuting groups of nonidentity four-qubit Pauli strings. Each group $\mathcal G_g$ contains $15$ strings. The corresponding setting applies a Clifford transformation $V_g$ such that every $V_gPV_g^\dagger$, with $P\in\mathcal G_g$, is a signed $Z$ parity. Computational-basis outcomes then give an eigenvalue $s_{gP}(b)\in\{-1,1\}$ for every $P$ in that group.

Let $r_g$ be the fraction of copies assigned to group $g$, with $\sum_g r_g=1$, and define the total coverage of $P$ by
\begin{equation}
t_P=\sum_{g:P\in\mathcal G_g}r_g.
\end{equation}
Methods for measuring Hamiltonians also redistribute measurable fragments to reduce sampling costs \cite{Choi2023FluidFragments}. We use every occurrence of a Pauli term, splitting each target into
\begin{equation}
H_{jg}=\sum_{P\in\mathcal G_g}\frac{r_g}{t_P}a_{jP}P,
\qquad O_j=\sum_g H_{jg}.
\label{eq:gc-splitting}
\end{equation}
For group $g$, the score is $\sum_{P\in\mathcal G_g}(r_g/t_P)a_{jP}s_{gP}(b)$. We assign $N_g=Nr_g$ copies to each group and add the sample means of these scores. The resulting variance constant is
\begin{equation}
\Gamma_{\rm GC}=\frac{d}{M(d+1)}
\sum_{P\ne I}\frac{e_P}{t_P}.
\label{eq:gc-risk}
\end{equation}
We optimize the nonnegative fractions $r_g$ over this fixed pool and omit groups with zero weight. We require $t_P>0$ for every term with $e_P>0$. The pool contains all maximal commuting Pauli groups, but it does not include arbitrary non-Clifford bases. In this comparison, we use this complete pool, include every occurrence of each Pauli term through Eq.~\eqref{eq:gc-splitting}, and optimize the allocation using Eq.~\eqref{eq:gc-risk}. Its value is $\Gamma_{\rm GC}=5.239637$.

\paragraph{Setting count within Clifford bases.}
\label{app:clifford-count}
The same four-qubit instance also separates the required numbers of
general and Clifford measurement bases. Every one of the $255$
nonidentity Pauli directions has a nonzero coefficient in at least one
target. An exact rational check on the stored targets gives
$\min_{P\ne I}\max_j|a_{jP}|>0.046$.
For the ideal targets of Appendix~\ref{app:example-certificate}, the
certified Frobenius perturbation below $10^{-11}$ changes each Pauli
coefficient by less than $2.5\times10^{-12}$, so this full support persists.
Each Clifford basis spans only $15$ nonidentity Pauli directions.
Exact reconstruction of the target space therefore requires
\begin{equation}
K_{\rm Clifford}\ge
\left\lceil\frac{255}{15}\right\rceil=17.
\label{eq:clifford-setting-count}
\end{equation}
The supplied Pauli spread partitions all $255$ directions into $17$
commuting groups of size $15$, so $17$ bases also suffice
\cite{Reggio2024Partitioning}. Thus $K_{\rm Clifford}=17$, whereas
the ideal construction has $K_\star=3$. This restriction refers to
Clifford bases in the original physical Pauli coordinates; it does not
include an arbitrary common non-Clifford pre-rotation. The count does
not assert that a $17$-basis allocation attains the minimum variance in
the full GC dictionary reported in Table~\ref{tab:comparison}.

\subsection{Globally rotated local joint measurement}
\label{app:joint}

We start from the local joint measurement of noisy Pauli observables in Ref.~\cite{McNulty2023Joint}. Unsharp measurements trade measurement sharpness for joint measurability \cite{Busch1986Unsharp,Busch2008Approximate}. On qubit $q$, its eight outcomes $(x,y,z)\in\{-1,1\}^3$ have effects
\begin{equation}
G_q(x,y,z)=\frac18
\bigl(I+\eta_{qx}xX+\eta_{qy}yY+\eta_{qz}zZ\bigr),
\label{eq:cube-povm}
\end{equation}
where $\eta_{qa}>0$ and $\sum_{a=x,y,z}\eta_{qa}^2=1$. We write $\beta_{qa}=\eta_{qa}^2$. In addition to these local strengths, we optimize a common global unitary $Q$ applied to the state before the local measurements. This global pre-rotation is our extension of the local protocol.

We implement the single-qubit POVM using four orthogonal projective measurement settings, as in Ref.~\cite{McNulty2023Joint}. For each pair $(\epsilon_y,\epsilon_z)\in\{-1,1\}^2$, we measure along the unit Bloch vector
\begin{equation}
\boldsymbol n_q=(\eta_{qx},\epsilon_y\eta_{qy},
\epsilon_z\eta_{qz}).
\end{equation}
If the observed eigenvalue is $s_q=\pm1$, we record $(x_q,y_q,z_q)=(s_q,s_q\epsilon_y,s_q\epsilon_z)$. We use the four settings with equal weights. Taking their tensor products gives $H=4^4=256$ settings, indexed by $h$. After applying $Q$, we measure $N/H$ copies in each product setting.

Expand the rotated targets as $QO_jQ^\dagger=\sum_{P\ne I}\widetilde a_{jP}P$, and set
\begin{equation}
\eta_P=\prod_{q:P_q\ne I}\eta_{q,P_q},
\qquad \beta_P=\eta_P^2.
\end{equation}
Let $s_P(h,b)$ be the product of the recorded $x_q$, $y_q$, or $z_q$ components specified by $P$. The standard joint-measurement score and estimator are
\begin{equation}
\begin{aligned}
g_j(h,b)&=\sum_{P\ne I}\frac{\widetilde a_{jP}}{\eta_P}s_P(h,b),\\
\widehat o_j&=\frac1H\sum_{h=1}^H\overline g_{jh},
\end{aligned}
\label{eq:joint-estimator}
\end{equation}
where $\overline g_{jh}$ averages this score over the copies assigned to setting $h$. For this fixed allocation, the variance constant is
\begin{equation}
\Gamma_{\rm joint}^{\rm fixed}
=\frac{d}{M(d+1)}\sum_{P\ne I}
\frac{\sum_j\widetilde a_{jP}^2}{\beta_P}.
\label{eq:joint-risk}
\end{equation}
We jointly optimize $Q$ and the positive strengths $\beta$, with equal sample allocation across all $256$ settings. The smallest value over nine initializations is $\Gamma_{\rm joint}^{\rm fixed}=5.633318$. The optimization is nonconvex, so this value need not be the global minimum.

This value depends on the fixed sample allocation. If we instead choose a setting independently and uniformly for each copy, use the same score, and average over all copies, we implement the same POVM on each copy, but the number of copies measured in each setting fluctuates. For the normalized traceless targets in this example,
\begin{equation}
\Gamma_{\rm joint}^{\rm iid}
=\frac{17}{16}\Gamma_{\rm joint}^{\rm fixed}-\frac1{51}
=5.965792.
\label{eq:joint-iid}
\end{equation}
These two values describe the standard score with equal allocation.

For the stronger comparison in Table~\ref{tab:comparison}, we allow both the readout and the setting fractions to vary. Let $R_h$ rotate the local axes of product setting $h$ to the computational axes. Its full measurement unitary is $U_h=R_hQ$. We expand a general centered outcome score in its measured parities,
\begin{equation}
f_{hj}(b)=\sum_{\varnothing\ne S\subseteq[n]}
      c_{hjS}\chi_S(b),
\label{eq:joint-general-readout}
\end{equation}
and use Eq.~\eqref{eq:unbiased-estimator}, with $N_h=Np_h$ and $\sum_h p_h=1$. We impose exact target reconstruction and optimize $C,p$ as in Eq.~\eqref{eq:fixed-basis-optimal-readout} for every fixed choice of $Q,\beta$. We also search over $Q$ and the local strengths. Unequal fractions change the mixture of product settings, so this enlarged protocol need not implement the equal-weight product POVM of Eq.~\eqref{eq:cube-povm}. It retains the same available local axes and global pre-rotation.

At the parameters optimized for equal allocation, optimizing the readout and fractions lowers the variance constant from $5.633318$ to $5.452793$. Allowing $Q,\beta$ to vary as well gives $\Gamma_{\rm joint}=4.796128$ reported in Table~\ref{tab:comparison}. The final readout and allocation solve is accurate at those measurement parameters; the nonconvex measurement search reaches its iteration limit, so we report an achieved value and do not claim full convergence or global optimality. For independent setting draws and the corresponding importance-weighted scores, the same protocol gives $\Gamma_{\rm joint}^{\rm iid}=5.076279$. The table uses fixed quotas.

\subsection{Global classical shadows}
\label{app:shadows}

Experimental demonstrations of classical-shadow estimation include Refs.~\cite{Struchalin2021ExperimentalShadows,Zhang2021ExperimentalShadows}.

For each copy, we draw $U$ independently from either the Haar measure on $U(d)$ or the uniform full $n$-qubit Clifford group, apply $U$, and measure the computational basis. From measurement outcome $b$, we construct the standard classical-shadow estimator \cite{Huang2020Shadows}
\begin{equation}
\widehat\rho=(d+1)U^\dagger|b\rangle\langle b|U-I.
\label{eq:shadow-snapshot}
\end{equation}
We use the same measurement data for every target and estimate its expectation value by the sample mean
\begin{equation}
\widehat o_j=\frac1N\sum_{t=1}^N
\operatorname{Tr}(O_j\widehat\rho_t).
\label{eq:shadow-estimator}
\end{equation}
We calculate the variance of this unbiased sample mean; we do not use a median-of-means estimator or a shadow-norm bound. The full multiqubit Clifford group is a unitary $3$-design \cite{Zhu2017Designs}, so its first two score moments agree with the global Haar protocol. Averaging over the pure input states gives
\begin{equation}
\begin{aligned}
\Gamma_{\rm shadow}
&=\frac{d+2}{d+1}\frac1M\sum_j\operatorname{Tr}(O_j^2)\\
&=\frac{96}{17}=5.647059.
\end{aligned}
\label{eq:shadow-risk}
\end{equation}
The random measurement unitaries specify the measurement protocol, while the Haar-random input states specify the common state average used to compare variances. We impose no restriction to local Pauli measurements.

%% file: six_qubit_protocols.tex
\subsection{Six-qubit construction and finite-copy comparison}
\label{app:six-qubit}

For Sec.~\ref{sec:six-qubit}, we fix the random seed to $20261018$. We form the $189\times96$ mixing matrix as in Eq.~\eqref{eq:mixing-matrix}, replacing the three block weights by $(0.7,1.1,1.6)$, each with block size $63$, and retaining the geometrically spaced singular values from $0.7$ to $1.3$. The left and right orthogonal factors have sizes $189\times96$ and $96\times96$. We normalize every target as in Eq.~\eqref{eq:six-targets}. Hardware connectivity can also constrain entangled measurement designs \cite{Escudero2023HardwareEntangled}.

Both the circuits used to generate the targets and the optimized circuits apply $R_z$ followed by $R_y$ on every qubit in each rotation block. Between blocks, CZ gates alternate between the disjoint pairs $\{(1,2),(3,4),(5,6)\}$ and $\{(2,3),(4,5),(6,1)\}$. Counting each parallel $R_z$, $R_y$, or CZ layer separately gives $3L-1$ gate layers before merging adjacent gates and performing the readout. Thus the generating circuits have $59$ gate layers and $57$ CZ gates, while the optimized $L=8$ circuits have $23$ gate layers and $21$ CZ gates. The two groups of random rotation angles are specified in Sec.~\ref{sec:six-qubit}.

We generate independent complex Gaussian Hermitian perturbations, make them traceless, subtract their Hilbert--Schmidt projections onto $\sum_r A(G_r)$, and normalize them to $\operatorname{Tr}(\Delta_j^2)/64=1/3$. Although these perturbations are orthogonal to the original measurement spaces, other measurement bases may still give an exact representation. The root mean squared target expectation under the Haar pure-state average is $1/\sqrt{195}$. Hence $\epsilon=0.1$ should not be interpreted as a $0.1$ floor on absolute RMSE.

We jointly optimize all three measurement unitaries at $N_0=10^4$. For fixed measurement unitaries and sample allocation, we solve the linear system for $C$ that minimizes Eq.~\eqref{eq:approximate-mse}; we also optimize the allocation fractions. We use several random initializations, initialize deeper circuits with solutions from shallower circuits, and use an initialization obtained from partial traces of the target operators. At each depth, we select the completed run with the lowest training MSE. We choose the smallest tested depth within $10\%$ of the best unperturbed training MSE, which gives $L=8$. For perturbed targets, we also allow initialization from the solution for the unperturbed targets. We do not use the unitaries used to generate the targets or their parameters in any of these steps. To calculate the curves, we fix each optimized set of measurement unitaries and solve the convex $C,p$ optimization problem for every $N$. The error floors in the plots are the least-squares projection errors in the limit of infinite $N$.

For any fixed collection of projective settings and allocation $p$, this finite-copy readout has a closed form. Use the real coordinates $\mathsf A_i,T$ and frame $F(p)$ of Eq.~\eqref{eq:fixed-basis-frame}, with inner product $\operatorname{Tr}(XY)/d$, and set $a=d/N$. Equation~\eqref{eq:approximate-mse} becomes
\begin{equation}
\begin{aligned}
M(d+1)\mathcal R_N
&=\left\|T-\sum_i\mathsf A_i C_i\right\|_F^2\\
&\quad+a\sum_i\frac{\|C_i\|_F^2}{p_i}.
\end{aligned}
\label{eq:finite-copy-coordinate-risk}
\end{equation}
At zero allocation we set $C_i=0$ and take its perspective term to be zero. For $N>0$, the unique solution on the positive-allocation blocks is
\begin{equation}
C_i^\star=p_i\mathsf A_i^{\mathsf T}
\left(F(p)+\frac dN I\right)^{-1}T,
\label{eq:finite-copy-readout}
\end{equation}
and its risk is
\begin{equation}
\min_C\mathcal R_N=\frac{d}{NM(d+1)}
\operatorname{Tr}\!\left[T^{\mathsf T}
\left(F(p)+\frac dN I\right)^{-1}T\right].
\label{eq:finite-copy-risk}
\end{equation}
To verify these expressions, let $X=(F+aI)^{-1}T$. The proposed coefficients give residual $T-FX=aX$ and satisfy the normal equations. The objective is strictly convex in the active coefficient blocks. Substitution gives $M(d+1)\mathcal R_N=a^2\operatorname{Tr}(X^{\mathsf T}X)+a\operatorname{Tr}(X^{\mathsf T}FX)=a\operatorname{Tr}(T^{\mathsf T}X)$. The shift $aI$ makes the inverse well defined even if $F$ is singular or the settings do not exactly cover the targets.

For fixed $p$, let $P_F$ project onto $\operatorname{ran}F$. The large-copy limits are
\begin{equation}
\begin{aligned}
\lim_{N\to\infty}\min_C\mathcal R_N
&=\frac{\|(I-P_F)T\|_F^2}{M(d+1)},\\
\lim_{N\to\infty}N\min_C\mathcal R_N
&=\frac{d}{M(d+1)}\operatorname{Tr}(T^{\mathsf T}F^+T),
\end{aligned}
\label{eq:finite-copy-risk-limits}
\end{equation}
where the second limit assumes $P_FT=T$ and recovers Eq.~\eqref{eq:fixed-basis-optimal-readout}. Formally, $N\to0^+$ gives $C_i^\star\to0$ and risk $\|T\|_F^2/[M(d+1)]$; this is a continuous-budget limit, not a prescription for fractional experimental counts. These formulas assume fixed quotas and do not apply unchanged to independently sampled shadow settings.

The comparison methods follow the preceding subsections with $n=6$, $d=64$, and $M=96$. The Pauli GC pool contains $987$ distinct groups: one fixed complete Pauli partition and three independently random-Clifford-rotated copies, together with all $3^6$ local Pauli settings, after removing duplicate groups. We optimize the sample allocation over this fixed pool, which does not contain all six-qubit Pauli groups. For the six-qubit joint measurement, we retain equal numbers of copies for the $4^6=4096$ settings and optimize the local strengths and a common global pre-rotation. This differs from the four-qubit table, where we also optimize the setting fractions and full linear readout. We parameterize it as $Q=U_L(\theta)U_L(0)^\dagger$, including the inverse of the zero-angle circuit so that zero parameters give the identity. We optimize the eight-block circuit from the identity and from random initializations, then use the best solution to initialize a twelve-block circuit. We perform at most $600$ optimization iterations at each stage. We therefore allow this comparison scheme to use a deeper pre-rotation than the shared circuits shown. As the optimization is nonconvex, we cannot guarantee a global minimum.

For local Pauli shadows~\cite{Huang2020Shadows}, we independently measure each qubit in a uniformly random $X$, $Y$, or $Z$ basis. The canonical score for a Pauli string $P$ is zero unless all nonidentity factors match the chosen axes; on a match, it is $3^{w(P)}$ times the product of measured eigenvalues, where $w(P)$ is the Pauli weight. We sum these scores with $a_{jP}$ to estimate $O_j$. The variance constant for target $j$, averaged over Haar-random pure states, is
\begin{equation}
\Gamma_{j,\mathrm{local}}
=\sum_{P\ne I}a_{jP}^2
\left(3^{w(P)}-\frac{1}{d+1}\right).
\label{eq:six-local-shadow}
\end{equation}

For spectral reuse, we pool all $96$ target eigenbases as in Appendix~\ref{app:spectral-reuse}. At each evaluated copy budget, we optimize the full finite-$N$ linear readout and the allocation for these fixed bases. We eliminate the readout with Eq.~\eqref{eq:finite-copy-readout} and optimize the resulting convex allocation objective using its analytic gradient. A direct Cholesky solve in the $96\times63=6048$ branch coordinates gives the readout and gradient. For Pauli GC, we likewise optimize the finite-$N$ allocation over the fixed $987$-group pool, together with its coordinatewise optimal linear readout. Appendix~\ref{app:baseline-convergence} describes the convergence checks and the resource choices used before calculating the curves. Lines for these two methods connect separately optimized budgets; no optimum is claimed at unevaluated budgets or for measurement bases outside the stated families.

For the Pauli-based protocols, we strengthen the finite-sample readout while retaining the measurement bases and joint-measurement parameters specified above. GC allocations vary with the copy budget as just described. Write $x_P=\langle P\rangle$ and, for a measured coordinate, let $\widehat x_P$ be the unbiased Pauli estimate supplied by the protocol. Averaging over Haar-random pure input states gives
\begin{equation}
\begin{aligned}
\mathbb E_\psi[x_Px_{P'}]&=\frac{\delta_{PP'}}{d+1},\\
\mathbb E_\psi[\operatorname{Cov}(\widehat x_P,\widehat x_{P'})]
&=\frac{v_P}{N}\delta_{PP'}.
\end{aligned}
\label{eq:six-pauli-covariance}
\end{equation}
For GC, the joint measurement, local shadows, and global shadows, respectively,
\begin{equation}
v_P=\begin{cases}
\displaystyle\frac{d}{(d+1)t_P},&\text{Pauli GC},\\[3pt]
\displaystyle\frac{d}{(d+1)\beta_P},&\text{rotated joint},\\[3pt]
3^{w(P)}-\displaystyle\frac1{d+1},&\text{local shadows},\\[3pt]
\displaystyle\frac{d(d+2)}{d+1},&\text{global shadows}.
\end{cases}
\label{eq:six-coordinate-noise}
\end{equation}
Here $t_P$ is the GC coverage at the allocation chosen for that copy budget. For the joint protocol, we use the coefficients of $QO_jQ^\dagger$ in place of those of $O_j$, and $\beta_P$ is the product of the optimized local strengths. Equation~\eqref{eq:six-pauli-covariance} concerns the state-averaged covariance; it does not assert diagonal covariance for each individual state.

We allow a separate shrinkage factor for every Pauli coordinate~\cite{Cai2025Biased}. For the estimator $\widehat o_j=\sum_{P\ne I}a_{jP}\lambda_P\widehat x_P$, the optimal factors and target-averaged MSE are
\begin{equation}
\begin{aligned}
\lambda_P^\star&=\frac{N}{N+(d+1)v_P},\\
\mathcal R_N^{\rm coord}
&=\frac1M\sum_{j,P\ne I}\frac{a_{jP}^2v_P}{N+(d+1)v_P}.
\end{aligned}
\label{eq:six-coordinate-shrinkage}
\end{equation}
For GC coordinates with $t_P=0$, we do not construct an unbiased Pauli
estimate. We set their readout to zero and interpret
Eqs.~\eqref{eq:six-coordinate-noise} and~\eqref{eq:six-coordinate-shrinkage}
by the continuous limit $t_P\to0^+$. Their contribution for target $j$
is then $a_{jP}^2/(d+1)$, arising entirely from bias.
This optimizes the readout within the family generated by the unbiased Pauli estimates. GC additionally optimizes its coverage through the allocation; the joint protocol retains the stated pre-rotation, strengths, and equal quotas. Global shadows have the same $v_P$ for every coordinate, so the optimum reduces to scalar shrinkage. Independent spectral measurements likewise retain their optimal per-target scalar shrinkage. If their unscaled variance is $\Gamma_j/N$ and $A_j=\operatorname{Tr}(O_j^2)/[d(d+1)]$, their risk is
\begin{equation}
\mathbb E_\psi[\operatorname{MSE}_j]
=\frac{A_j\Gamma_j}{NA_j+\Gamma_j}.
\label{eq:six-scalar-shrinkage}
\end{equation}
We average the MSE over targets before taking the square root. All fixed allocations use continuous fractions. In particular, equal integer counts for the joint protocol require $N$ to be a multiple of $4096$; its low-$N$ curve is an idealized comparison favorable to that protocol. All schemes assume ideal gates, ideal readout, and independent copies.

%% file: baseline_convergence.tex
\subsection{Convergence of the comparison allocations}
\label{app:baseline-convergence}

We assess convergence of the allocation optimization for spectral reuse and Pauli GC in the
six-qubit comparison. We keep the measurement bases fixed: spectral reuse
uses the same $96$ target eigenbases, and Pauli GC uses the same $987$-group
dictionary. At each evaluated total copy budget $N$, we minimize
$f_N(p)=\min_C\mathcal R_N(C,p)$ over the allocation simplex.
Equation~\eqref{eq:finite-copy-coordinate-risk} is jointly convex in $C,p$,
so $f_N$ is convex in $p$. Spectral reuse uses the full linear readout.
For Pauli GC the frame is diagonal in Pauli coordinates, so the
coordinatewise readout in Eq.~\eqref{eq:six-coordinate-shrinkage} is also
the full linear optimum for the fixed dictionary. Writing
$\bar e_P=e_P/M=M^{-1}\sum_j a_{jP}^2$, its allocation objective is
\begin{equation}
f_N^{\rm GC}(p)=\sum_{P\ne I}
\frac{d\,\bar e_P}{(d+1)[N(Bp)_P+d]},
\label{eq:gc-finite-allocation}
\end{equation}
where $B_{Pi}=1$ when setting $i$ measures $P$ and is zero otherwise.

\begin{figure*}[t]
\centering
\includegraphics[width=0.94\textwidth]{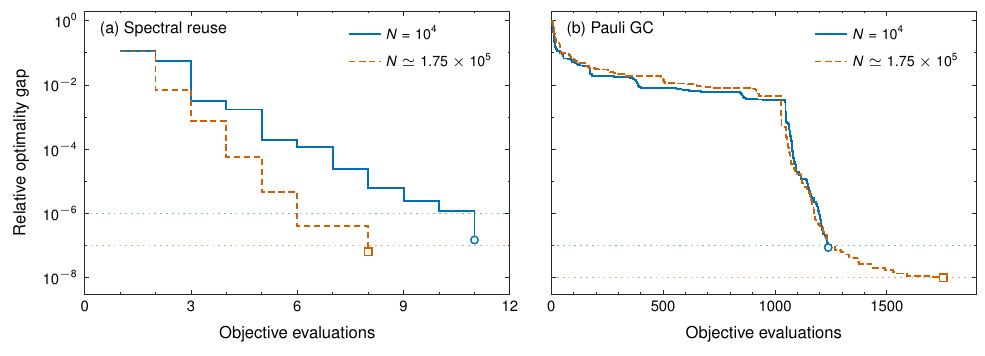}
\caption{Allocation convergence for the perturbed six-qubit targets
($\epsilon=0.1$): (a) spectral reuse and (b) Pauli GC. Both use fixed
measurement bases and uniform initial allocation. The copy budgets are
$N=10^4$ and $N=175192$; the latter is rounded in the legend.
The vertical axis shows $\eta=(U-L)/U$, with $U$ the best feasible MSE
and $L$ the accumulated first-order lower bound. The horizontal axis
counts joint objective/gradient evaluations, including line-search and
infeasible trial calls. Thin horizontal dotted lines show the selected
tolerances in Table~\ref{tab:allocation-tolerances}. Open markers identify
their first attainment, where the displayed traces end. The checks concern
the allocation objective within each fixed measurement family; evaluation costs
differ between methods.}
\label{fig:allocation-convergence}
\end{figure*}

A nearly constant MSE alone does not establish optimization accuracy.
At a feasible allocation, the convex first-order gap is
\begin{equation}
g_N(p)=p^{\mathsf T}\nabla f_N(p)-\min_i[\nabla f_N(p)]_i.
\end{equation}
The tangent inequality gives
\begin{equation}
f_N(p)-g_N(p)\le f_N^\star\le f_N(p),
\label{eq:allocation-convex-gap}
\end{equation}
where $f_N^\star$ is the minimum for these fixed bases. For convergence
diagnostics, we retain the smallest feasible value $U$ and the largest
valid tangent lower bound $L$ encountered, starting with the lower bound
$L=0$. The accumulated relative gap $\eta=(U-L)/U$ then bounds the
remaining improvement relative to the current best MSE:
$0\le U-f_N^\star\le\eta U$. A decrease in $\eta$ can reflect a tighter
lower bound even when the MSE changes little.

GC uses the accumulated gap as its stopping criterion. Spectral reuse
tests $g_N(p)/f_N(p)$ at the accepted allocation; the accumulated gap
reported in the figure can be smaller. Only feasible trial points
contribute to the upper and lower bounds. All gaps are numerical checks
subject to floating-point and linear-solve accuracy, not interval-arithmetic
certificates.

We use pilot runs at representative budgets to choose the tolerances in
Table~\ref{tab:allocation-tolerances} and the protective iteration limits
before computing the comparison curves. Figure~\ref{fig:allocation-convergence}
shows two of these uniform-initialization checks for the perturbed targets:
$N=10^4$ and $N=175192$, near the observed crossover. Here $N$ counts
quantum-state copies, while the horizontal axis counts classical
objective evaluations. Each call evaluates both the objective and its
gradient and counts once, including line-search and infeasible trial calls.
Basis preparation and separate final validation are excluded from this count.
The figure displays each trajectory through its first attainment of the
selected production tolerance; the original GC pilot runs continued beyond
these points. Evaluation costs differ between the methods, so the two
horizontal scales do not provide a runtime comparison.

Production may reuse a validated pilot solution or initialize from a
neighboring budget; these choices are recorded with each point.
The protective limits are $200$ objective/gradient evaluations per
spectral-reuse point, with a $900$ s guard, and $1000$ L-BFGS,
$14153$ accelerated projected-gradient, and $20000$ Frank--Wolfe iterations
for GC. Every evaluated spectral-reuse and GC budget used in
Fig.~\ref{fig:six-qubit-results} passes its numerical gap test before
these limits; acceptance does not rely on the optimizer's success flag.

\begin{table}[t]
\caption{Relative allocation-gap upper limits for the six-qubit curves.
The crossover neighborhood is $1.4\times10^5\le N\le2.1\times10^5$;
some GC points outside it use a tighter tolerance as well.
The measurement bases remain fixed in both optimizations.}
\label{tab:allocation-tolerances}
\begin{ruledtabular}
\begin{tabular}{lrr}
Method & Other budgets & Crossover region\\
\hline
Spectral reuse & $10^{-6}$ & $10^{-7}$\\
Pauli GC & $10^{-7}$ & $10^{-8}$\\
\end{tabular}
\end{ruledtabular}
\end{table}

The sources retain the full pilot histories and the allocation and gap at
every evaluated comparison budget. These checks bound the remaining
reduction in MSE within the stated measurement families. They do not
optimize the choice of target eigenbases or the Pauli group dictionary,
and do not establish optimality over all projective measurements.

For the perturbed instance, independent solves at $N=175190$ and $175193$
give opposite orderings between the shared design and spectral reuse,
even after including both numerical allocation-gap margins. With the
learned shared unitaries fixed, both endpoint allocations attain relative
first-order gaps below $4\times10^{-10}$, corresponding to MSE gaps below
$2.5\times10^{-14}$. A local
estimate gives $N_\times\simeq175191.05$ and
$\mathrm{RMSE}(N_\times)\simeq0.00803714$, consistent with the rounded
values in Eq.~\eqref{eq:six-crossing}. This is the observed crossover of
the specified designs; the numerical bracket does not establish an
analytic unique-crossing theorem.

%% file: initialization_checks.tex
\subsection{Additional initialization checks}
\label{app:initialization-checks}

Separate checks on the same target matrices illustrate the sensitivity of
the nonconvex search. These checks use independent implementations and
bounded runs; they are not a replay of the training used for
Table~\ref{tab:comparison} or Fig.~\ref{fig:six-qubit-results}.

For the four-qubit targets, we tested five specified independent Haar
initializations with $K=3$, allowing at most $1000$ objective evaluations
or $120$ seconds per run. The final joint relative Frobenius residuals
were $0.214$, $1.35\times10^{-11}$, $7.68\times10^{-7}$,
$1.16\times10^{-11}$, and $0.226$, after $1000$, $418$, $1000$,
$432$, and $364$ evaluations, respectively. Two runs reached the stated
threshold of $10^{-8}$. The other stopping reasons were the evaluation
limit for the first and third runs, and a floating-point function-improvement
criterion for the fifth. These runs minimize reconstruction error, without
optimizing or recording their variance constants. They therefore do not
establish finite-sample performance or a general success probability.
A single evaluation of the archived solution gave a residual of
$2.61\times10^{-14}$; this was an implementation check, with no further
optimization.

For the clean six-qubit targets at $K=3$, $L=8$, and $N=10^4$, a
$300$-iteration continuation of the archived solution reduced the risk
from $9.57852\times10^{-5}$ to $9.57670\times10^{-5}$, an improvement
of about $0.019\%$. Two independent random starts ended at
$4.20038\times10^{-4}$ and $8.93419\times10^{-4}$. For the perturbed
targets, a continuation reduced the risk from $1.45549083\times10^{-4}$
to $1.45548347\times10^{-4}$. All four runs reached the $300$-iteration
limit. These finite-budget checks do not establish convergence.
The source archive contains the run records, target identifiers, and a
script that checks their consistency.

%% file: structural_capacity_proof.tex
\subsection{Structural lower bound}
\label{app:proof-structural}

\begin{proof}[Proof of Theorem~\ref{thm:structural-capacity}]
Write $r=d-1$ and $\mathcal A_i=A(U_i)$, and suppose that
$V\subseteq S=\sum_{i=1}^K\mathcal A_i$. Set
\begin{equation}
D=\dim_{\mathbb R}S,\qquad
G=S\cap F,\qquad b=\dim_{\mathbb R}G.
\end{equation}
Each $\mathcal A_i$ has dimension $r$, so $D\le Kr$. Since
$V\subseteq S$, we also have $V\cap G=V\cap F$ and $b\ge h$.
Moreover, $V+G\subseteq S\cap\mathcal W$, which gives
\begin{equation}
\dim_{\mathbb R}(S\cap\mathcal W)\ge m+b-h.
\label{eq:structural-proof-intersection}
\end{equation}
Define $\delta=D-m-b+h\ge0$, the codimension of $V+G$ in $S$.
For each $i$, the dimension inequality for intersections in $S$
then gives
\begin{equation}
\begin{aligned}
\dim_{\mathbb R}(\mathcal A_i\cap\mathcal W)
&\ge r+\dim_{\mathbb R}(S\cap\mathcal W)-D\\
&\ge r-\delta.
\end{aligned}
\label{eq:structural-proof-li}
\end{equation}

Let $g_i=\dim_{\mathbb R}(\mathcal A_i\cap F)$. The space
$\mathcal A_i\cap\mathcal W$ is internally commuting. Every
element of $F=Z(\mathcal W)$ commutes with all of $\mathcal W$,
so $(\mathcal A_i\cap\mathcal W)+F$ is a commuting subspace of
$\mathcal W$. Consequently,
\begin{equation}
\dim_{\mathbb R}(\mathcal A_i\cap\mathcal W)+c-g_i
\le\alpha(\mathcal W).
\end{equation}
Together with Eq.~\eqref{eq:structural-proof-li}, this implies
\begin{equation}
g_i\ge r-\delta+c-\alpha(\mathcal W).
\label{eq:structural-proof-gi}
\end{equation}

Choose a linear complement $E_i$ such that
$\mathcal A_i=(\mathcal A_i\cap F)\oplus E_i$ and
$\dim_{\mathbb R}E_i=r-g_i$. Since
$\mathcal A_i\cap F\subseteq G\subseteq S$, the space
$G+\sum_iE_i$ equals $S$. Hence
\begin{equation}
\begin{aligned}
D&\le b+\sum_{i=1}^K(r-g_i)\\
&\le b+K\bigl[\delta+\alpha(\mathcal W)-c\bigr].
\end{aligned}
\end{equation}
We do not assume here that
$G=\sum_i(\mathcal A_i\cap F)$. Substituting the definition of
$\delta$ and rearranging gives
\begin{equation}
\begin{aligned}
Km+Kc-K\alpha(\mathcal W)
 +(K-1)b-Kh\le(K-1)D.
\end{aligned}
\end{equation}
Using $b\ge h$, $D\le Kr$, and $K-1\ge0$, we obtain
\begin{equation}
Km+Kc-K\alpha(\mathcal W)-h\le K(K-1)r.
\end{equation}
Dividing by $K$ yields
\begin{equation}
m\le(K-1)r+\alpha(\mathcal W)-c+\frac hK.
\end{equation}
All terms except $h/K$ are integers, so taking its floor proves
Eq.~\eqref{eq:structural-bound}. The proof includes $K=1$, for
which the condition becomes $m+c-h\le\alpha(\mathcal W)$.
\end{proof}

We used only the inclusion $V\subseteq S$. The measurement
spaces may contain directions outside $\mathcal W$ or $V$, and
neither of these spaces needs to be closed under products or
commutators.

The common-symmetry bound follows by choosing
\begin{equation}
\mathcal W=\{X\in\operatorname{Herm}_0(d):
[X,Q_\mu]=0\text{ for all }\mu\},
\end{equation}
where the Hermitian $Q_\mu$ commute with every target but need not
commute with one another. To evaluate these parameters, consider the block
decomposition
\begin{equation}
W_Q^\dagger\mathfrak C_QW_Q
=\bigoplus_{\lambda=1}^s
\bigl(I_{u_\lambda}\otimes M_{v_\lambda}(\mathbb C)\bigr),
\end{equation}
where
$\mathfrak C_Q=\{X\in M_d(\mathbb C):[Q_\mu,X]=0\ \forall\mu\}$
and $\sum_\lambda u_\lambda v_\lambda=d$. Then
\begin{equation}
\alpha(\mathcal W)=\sum_{\lambda=1}^s v_\lambda-1,
\qquad c=s-1.
\end{equation}
A commuting Hermitian family can be diagonalized within each
matrix block, leaving $\sum_\lambda v_\lambda$ real parameters
before the global trace constraint. The center has one scalar
per block, subject to the same constraint. We obtain the remaining parameter
$h=\dim_{\mathbb R}(V\cap F)$ by a linear rank calculation.
This recovers the common-commutant formula.

Finally, let $L=\lceil m/r\rceil$. For every $K\ge L+1$, the
right-hand side of Eq.~\eqref{eq:structural-bound} is at least
$(K-1)r\ge Lr\ge m$, because $\alpha(\mathcal W)\ge c$ and
$h\ge0$. Thus this theorem alone can improve the dimension
lower bound by at most one setting.

%% file: witness_proofs.tex
\subsection{Odd alternating traces for two settings}
\label{app:proof-odd-trace}

\begin{proof}[Proof of Theorem~\ref{thm:odd-trace-witness}]
Suppose $V\subseteq\mathcal A+\mathcal B$, where each of $\mathcal A$ and $\mathcal B$ is internally commuting. For arbitrary $X_1,\ldots,X_q\in V$, choose decompositions $X_j=A_j+B_j$, with $A_j\in\mathcal A$ and $B_j\in\mathcal B$. By multilinearity, we expand $\tau_q$ into terms in which each labeled input belongs to a fixed family, $\mathcal A$ or $\mathcal B$.

Fix one such choice. For each permutation $\pi$, consider its family labels in cyclic order. An odd cycle with two labels must have an edge whose endpoints have the same label. Otherwise the labels would alternate around the cycle, which is impossible for odd $q$.

Number the cyclic edges by $(1,2),\ldots,(q-1,q),(q,1)$. Choose the first equal-label edge and swap the two input labels at its endpoints. Their family labels agree, so the entire cyclic family pattern stays unchanged. The same edge is therefore selected after the swap. This rule pairs permutations without fixed points and reverses their signs.

The two paired traces are equal. For an ordinary adjacent edge, the swapped factors commute because they belong to the same family. For the edge $(q,1)$, cyclicity first brings the endpoint factors together; commutation again gives equal traces. More explicitly, if the ordered factors are $Y_1,\ldots,Y_q$ and $[Y_1,Y_q]=0$, then
\begin{equation}
\begin{aligned}
\operatorname{Tr}(Y_qY_2\cdots Y_{q-1}Y_1)
&=\operatorname{Tr}(Y_1Y_qY_2\cdots Y_{q-1})\\
&=\operatorname{Tr}(Y_qY_1Y_2\cdots Y_{q-1})\\
&=\operatorname{Tr}(Y_1Y_2\cdots Y_q).
\end{aligned}
\end{equation}
Each pair cancels in the signed sum. Hence the alternating trace is zero for the fixed family assignment. Summing over all assignments proves the theorem. Two measurement settings provide two internally commuting spaces, so the result applies to the measurement model of Sec.~\ref{sec:framework}.
\end{proof}

For $q=3$, cyclicity groups the six permutations into two sets of three and gives
\begin{equation}
\begin{aligned}
\tau_3(X,Y,Z)
&=\tfrac12\bigl[\operatorname{Tr}(XYZ)-\operatorname{Tr}(XZY)\bigr]\\
&=\tfrac12\operatorname{Tr}\!\left(X[Y,Z]\right).
\end{aligned}
\end{equation}
This gives the normalization in Eq.~\eqref{eq:cubic-witness}. For Hermitian inputs, $\operatorname{Tr}(X[Y,Z])$ is purely imaginary, so $\Omega_0$ is real. Subtracting an identity component from any input does not change $\Omega_0$, because the identity commutes with every operator and the trace of a commutator is zero.

\subsection{Certificate for the four-qubit example}
\label{app:example-certificate}

We distinguish the exact constructed instance from its archived floating-point target matrices $T_j$. Let $\widehat G_r$ be the archived generating matrices and interpret the entries of $W$ as exact binary rationals. We define each exact generating unitary by its unitary polar factor,
\begin{equation}
G_r=\widehat G_r
\bigl(\widehat G_r^\dagger\widehat G_r\bigr)^{-1/2},
\end{equation}
and use Eq.~\eqref{eq:example-targets} with exact normalization to define $O_j$. These targets have an exact representation using three settings by construction. The source archive includes the finite inputs and the scripts for the following rational-arithmetic checks.

First, for every $r$, the checked Gram defect satisfies
\begin{equation}
\|\widehat G_r^\dagger\widehat G_r-I\|_F<\delta,
\qquad \delta=10^{-14}.
\end{equation}
The generating matrices are therefore invertible. Their singular values give $\|G_r-\widehat G_r\|_F<\delta$ and $\|\widehat G_r\|_{\rm op}<2$. Hence every Pauli parity obeys
\begin{equation}
\|G_r^\dagger Z_SG_r-
\widehat G_r^\dagger Z_S\widehat G_r\|_F<3\delta.
\end{equation}

For each target, choose a positive rational scale $\alpha_j$ and form the exact rational matrix
\begin{equation}
\widehat H_j=\alpha_j\sum_{r,S}w_{rSj}
\widehat G_r^\dagger Z_S\widehat G_r.
\end{equation}
The archived scales satisfy $\sum_{r,S}|\alpha_jw_{rSj}|<4$, $\|\widehat H_j-T_j\|_F<10^{-14}$, and $|\|T_j\|_F^2-16/3|<10^{-13}$. All these comparisons use exact rational numbers. Replacing $\widehat G_r$ by $G_r$ changes this matrix by less than $12\delta$. Normalizing it to Frobenius norm $4/\sqrt3$ then gives
\begin{equation}
\|O_j-T_j\|_F
<2(12\delta+10^{-14})+5\times10^{-14}
<10^{-11}.
\label{eq:certified-target-distance}
\end{equation}
The positive scale cancels in the exact normalization.

Second, exact evaluation of the archived matrices gives
\begin{equation}
\begin{aligned}
-0.025089497
&<\Omega_0(T_{15},T_{16},T_{28})\\
&<-0.025089496.
\end{aligned}
\end{equation}
Both the exact and archived target norms are less than $3$. Multilinearity and $|\operatorname{Tr}(ABC)|\le\|A\|_F\|B\|_F\|C\|_F$ bound the change of this cubic value by
\begin{equation}
|\Delta\Omega_0|<\frac{6\cdot3^2}{16}\,10^{-11}
=3.375\times10^{-11}.
\end{equation}
It follows that the exact constructed targets satisfy
\begin{equation}
-0.025090<\Omega_0(O_{15},O_{16},O_{28})<-0.025089.
\label{eq:certified-witness-interval}
\end{equation}
Theorem~\ref{thm:odd-trace-witness} thus excludes two settings, while our construction gives an exact representation with three. This proves Eq.~\eqref{eq:example-minimum}. We use the archived matrices to optimize the bases and compare variances, and report the reconstruction accuracy separately in Eq.~\eqref{eq:example-residual}.

%% file: four_device_example.tex
\subsection{An exact example with four measurement settings}
\label{app:four-device-example}

We give the construction and statistical details for the collective-moment
task introduced in Sec.~\ref{sec:collective-moments}. The targets are
defined in Eq.~\eqref{eq:collective-targets}; we order them as
\begin{equation}
\mathcal O=\{J_x,J_y,J_z,Q_{xx},Q_{yy},Q_{zz},
Q_{xy},Q_{xz},Q_{yz}\},
\end{equation}
with an arbitrary input state on the full $d=4$ space.
Their Pauli expansions have disjoint support, so they are
nonzero and pairwise Hilbert--Schmidt orthogonal. Their real
span $V$ therefore has dimension $m=9$.

We use the triplet--singlet basis
\begin{equation}
\begin{aligned}
|t_0\rangle&=|00\rangle,&
|t_1\rangle&=(|01\rangle+|10\rangle)/\sqrt2,\\
|t_2\rangle&=|11\rangle,&
|s\rangle&=(|01\rangle-|10\rangle)/\sqrt2,
\end{aligned}
\end{equation}
and denote the associated sector projectors by $P_t$ and $P_s$.
All targets commute with swap. They thus belong to the real space
\begin{equation}
\begin{aligned}
\mathcal W=\{B\oplus b:\;&B=B^\dagger\in M_3(\mathbb C),\quad
b\in\mathbb R,\\
&\operatorname{Tr}B+b=0\}.
\end{aligned}
\end{equation}
This is the traceless Hermitian part of the block algebra
$M_3(\mathbb C)\oplus M_1(\mathbb C)$. Since
$\dim_{\mathbb R}\mathcal W=3^2+1^2-1=9$ and
$V\subseteq\mathcal W$, we have $V=\mathcal W$.

The internal center is $F=\mathbb R H_{\mathrm c}$, with
\begin{equation}
\begin{aligned}
H_{\mathrm c}
&=J_x^2+J_y^2+J_z^2-\tfrac32I_4\\
&=\tfrac12P_t-\tfrac32P_s
=Q_{xx}+Q_{yy}+Q_{zz}.
\end{aligned}
\end{equation}
Indeed, a Hermitian operator commuting with every Hermitian
matrix on the triplet block must be scalar on that block; the
global trace constraint then leaves one central direction.
Thus $c=h=1$. A commuting family in $\mathcal W$ can be
diagonalized within the triplet block, giving at most three
independent traceless directions overall. The commuting,
independent operators $H_{\mathrm c},J_z,Q_{zz}$ attain this
dimension, so $\alpha(\mathcal W)=3$.

If three arbitrary global measurement bases covered $V$,
Eq.~\eqref{eq:structural-bound} would require
\begin{equation}
9\le(3-1)\cdot3+3-1+\lfloor1/3\rfloor=8.
\end{equation}
Hence $K_\star(V)\ge4$, whereas dimension counting alone gives
only $K_\star(V)\ge\lceil9/3\rceil=3$. The center correction
is essential here: omitting it would give $9\le9$.

We next construct four bases attaining this bound, using a complete set of mutually unbiased bases in the triplet sector \cite{Wootters1989MUB}. Let
$\omega=e^{2\pi i/3}$ and take the computational triplet basis
$\mathcal T_\infty=\{|t_0\rangle,|t_1\rangle,|t_2\rangle\}$.
For $a=0,1,2$, define
\begin{equation}
|\psi_{a,b}\rangle=\frac1{\sqrt3}\sum_{x=0}^2
\omega^{ax^2+bx}|t_x\rangle,\qquad b=0,1,2.
\label{eq:four-device-chirps}
\end{equation}
Write $|\psi_{\infty,b}\rangle=|t_b\rangle$ and
$\mathcal T_\mu=\{|\psi_{\mu,b}\rangle\}_{b=0}^2$, where
$\mu\in\{\infty,0,1,2\}$. Each $\mathcal T_a$ is orthonormal
by the finite Fourier identity. Its vectors have squared overlap
$1/3$ with each computational triplet vector. For $a\ne a'$,
put $A=a'-a\ne0$ and $B=b'-b$, with arithmetic modulo three.
The remaining overlaps obey
\begin{equation}
\begin{aligned}
\left|\sum_{x\in\mathbb F_3}\omega^{Ax^2+Bx}\right|^2
&=\sum_{u\in\mathbb F_3}\omega^{Au^2+Bu}
  \sum_{y\in\mathbb F_3}\omega^{2Auy}\\
&=3,
\end{aligned}
\end{equation}
because only $u=0$ survives the inner sum. Thus the four triplet
bases are mutually unbiased:
\begin{equation}
|\langle\psi_{\mu,b}|\psi_{\nu,c}\rangle|^2=\tfrac13,
\qquad\mu\ne\nu.
\end{equation}

Within the triplet sector, the union of these bases forms a complex projective 2-design \cite{Klappenecker2005Designs}. We add the singlet to obtain four complete measurement bases,
\begin{equation}
\mathcal B_\mu=
\{|\psi_{\mu,0}\rangle,|\psi_{\mu,1}\rangle,
|\psi_{\mu,2}\rangle,|s\rangle\}.
\end{equation}
If $B_\mu$ has these basis vectors as columns, the convention
$A(U)=U^\dagger\mathcal D_0U$ assigns the measurement unitary
$U_\mu=B_\mu^\dagger$. The full four-dimensional bases share
$|s\rangle$ and are not mutually unbiased; mutual unbiasedness
holds only in the three-dimensional triplet sector.

To verify coverage, let
\begin{equation}
\begin{aligned}
E_{\mu,b}&=|\psi_{\mu,b}\rangle\langle\psi_{\mu,b}|-\tfrac13P_t,\\
\mathcal L_\mu&=\operatorname{span}_{\mathbb R}\{E_{\mu,b}\}_{b=0}^2.
\end{aligned}
\end{equation}
For each $\mu$, $\sum_bE_{\mu,b}=0$ and
$\dim_{\mathbb R}\mathcal L_\mu=2$. Mutual unbiasedness implies
\begin{equation}
\operatorname{Tr}(E_{\mu,b}E_{\nu,c})
=|\langle\psi_{\mu,b}|\psi_{\nu,c}\rangle|^2-\tfrac13=0
\end{equation}
when $\mu\ne\nu$. These four orthogonal two-dimensional spaces
therefore span all eight traceless triplet directions. Each full
measurement space satisfies
\begin{equation}
A(U_\mu)=\mathcal L_\mu\oplus\mathbb R H_{\mathrm c},
\end{equation}
so
\begin{equation}
\sum_\mu A(U_\mu)
=\left(\bigoplus_\mu\mathcal L_\mu\right)
\oplus\mathbb R H_{\mathrm c}
=\mathcal W=V.
\end{equation}
The four bases give $K_\star(V)\le4$. Combining the upper and lower bounds, we obtain
\begin{equation}
K_\star(V)=4.
\label{eq:four-device-exact-minimum}
\end{equation}
The complete measurement spaces overlap in the central direction
$H_{\mathrm c}$; only their triplet-traceless parts are mutually
orthogonal. We imposed no restriction on the input state. We allow singlet
population and triplet--singlet coherences; the latter do not
contribute to the expectation values of these block-diagonal
targets.

\subsubsection{Statistical comparison}
\label{app:collective-statistics}

We compare precision for the same nine targets with their natural
normalization. Their squared Hilbert--Schmidt norms, in the order above, are
\begin{equation}
\bigl(\operatorname{Tr}O_j^2\bigr)_{j=1}^9
=(2,2,2,1,1,1,\tfrac12,\tfrac12,\tfrac12).
\end{equation}
We use the target-average Haar pure-state variance $\Gamma/N$
of Eq.~\eqref{eq:variance-constant}, on the full $d=4$ space.
Allocations are fixed except for the independent global-shadow
draws \cite{Huang2020Shadows}.

\begin{table}[tbp]
\caption{Variance for the nine natural collective moments.
The second column counts distinct measurement bases for the
specified implementations. Independent spectral measurements
use nine separate target data sets in six distinct bases.
}
\label{tab:collective-comparison}
\begin{ruledtabular}
\begin{tabular}{lrr}
Scheme & Bases & $\Gamma$\\
\hline
Optimized shared bases & $4$ & $0.799782$\\
Triplet MUBs and singlet & $4$ & $0.847364$\\
Spectral reuse & $6$ & $1.011111$\\
Pauli GC & $5$ & $1.066667$\\
Global Haar/Clifford shadows & --- & $1.400000$\\
Independent spectral & $6$ & $1.948528$\\
\end{tabular}
\end{ruledtabular}
\end{table}

Optimizing the readout and allocation of the four explicit bases gives
$\Gamma=0.847364$; equal allocation gives $0.866667$.
We then vary all four triplet bases, keep the singlet fixed, and
optimize the readout and allocation. The explicit construction
and two independent random initializations yield $\Gamma=0.799782$.
The selected solution reconstructs the targets with maximum relative
Frobenius error $6.72\times10^{-16}$. Its measurement space equals
$\mathcal W$, so it still uses the minimum number of measurement settings.
This achieved variance is not a certified global optimum.
The sources include the bases, coefficients, and allocations.

The protocols follow Appendix~\ref{app:protocols}, with $d=4$ and $M=9$.
Spectral reuse combines all target eigenbases with optimized readout and allocation
\cite{Watrous2018,DAriano2007Optimal,Fischer2024DualFrame}.
We keep the singlet fixed and use the $J_\alpha$ eigenbasis
for both $J_\alpha$ and $Q_{\alpha\alpha}$, giving six distinct bases.
The optimized fractions are $2/9$ for each common axis basis
and $1/9$ for each cross-moment basis. We fix the choices
within degenerate eigenspaces.

For Pauli GC, we optimize over all fifteen maximal commuting
two-qubit Pauli groups \cite{Yen2020Compatible}. The optimal
value in this dictionary is $16/15$, and five groups suffice
to attain it. To verify optimality, use
$e_P=\sum_j a_{jP}^2$ and $\sum_P t_P=3$ in
Eq.~\eqref{eq:gc-risk}. Cauchy's inequality gives
\begin{equation}
\Gamma_{\rm GC}\ge
\frac4{135}\left(\sum_{P\ne I}\sqrt{e_P}\right)^2
=\frac{16}{15}.
\end{equation}
The supplied allocation attains this bound. All fifteen
nonidentity Pauli directions occur, and each Pauli basis covers
three, so at least five such bases are necessary. Our four
shared bases need not be Clifford bases.

Spectral reuse and Pauli GC require about $1.26$ and $1.33$
times as many copies, respectively, for the same mean variance.
This modest statistical gain accompanies an exact certificate for the minimum number of settings. The comparison concerns averages, rather than
every state or target separately.

\input{collective_variance_bound}

%% file: collective_variance_bound.tex
\subsubsection{A variance bound for arbitrary global bases}
\label{app:collective-global-bound}

For the nine targets with the natural normalization above, every exact
unbiased linear reconstruction from a finite collection of projective
measurements on the original system obeys
\begin{equation}
\Gamma\ge\Gamma_{\rm LB}
=\frac{31}{60}+\frac{\sqrt2}{5}
=0.7995093791\ldots .
\label{eq:collective-pvm-lower-bound}
\end{equation}
We use fixed allocations, independent copies, equal target weights, and
the Haar pure-state average on the full four-dimensional space. The
bound allows any number of settings and arbitrary global measurement
bases, including bases that mix the triplet and singlet sectors.

\paragraph{Proof.}
Use the real operator inner product $(X,Y)_0=\operatorname{Tr}(XY)/4$,
with norm $\|X\|_0^2=(X,X)_0$. Let $P_{\mathcal A_i}$ be the
orthogonal projector in operator space onto the $i$th measurable
traceless space. The frame of Eq.~\eqref{eq:fixed-basis-frame} and
the target weight operator are
\begin{equation}
F=\sum_i p_iP_{\mathcal A_i},\qquad
\mathsf S=\sum_{j=1}^9|O_j)(O_j|.
\end{equation}
Here $|X)(Y|$ maps $Z$ to $X(Y,Z)_0$, and
$0\preceq F\preceq I_{\rm op}$. Exact reconstruction requires
$V\subseteq\operatorname{ran}F$. Constant terms in the individual
setting scores can be removed without changing their variance; their
sum vanishes for the traceless targets. We write $\operatorname{tr}_{\rm op}$ for the trace of a linear map on
the real space of traceless Hermitian operators. The fixed-frame optimum in
Eq.~\eqref{eq:fixed-basis-optimal-readout} becomes
\begin{equation}
\Gamma_{\min C}=\frac4{45}
\operatorname{tr}_{\rm op}(\mathsf S F^+).
\label{eq:collective-frame-optimum}
\end{equation}
This also follows by completing the square. For any decomposition
$O_j=\sum_iD_{ij}$ with $D_{ij}\in\mathcal A_i$, set $Y_j=F^+O_j$
and omit zero-allocation settings. Then
\begin{equation}
\begin{aligned}
\sum_i\frac{\|D_{ij}-p_iP_{\mathcal A_i}Y_j\|_0^2}{p_i}
&=\sum_i\frac{\|D_{ij}\|_0^2}{p_i}\\
&\quad -(O_j,F^+O_j)_0\ge0.
\end{aligned}
\end{equation}
The choice $D_{ij}=p_iP_{\mathcal A_i}F^+O_j$ attains equality
and reconstructs $O_j$. Thus any feasible readout has variance at
least Eq.~\eqref{eq:collective-frame-optimum}.

Write
\begin{equation}
h_0=\frac{XX+YY+ZZ}{\sqrt3}
=\frac{P_t-3P_s}{\sqrt3},\qquad (h_0,h_0)_0=1,
\end{equation}
where $XX=\sigma_x\otimes\sigma_x$, and similarly for $YY,ZZ$.
Let $\mathcal T$ be the eight-dimensional space of traceless
Hermitian operators supported on the triplet sector. We have
$V=\mathbb Rh_0\oplus\mathcal T$ and
\begin{equation}
\mathsf S=\tfrac14|h_0)(h_0|+\mathsf S_{\mathcal T},\qquad
\operatorname{spec}(\mathsf S_{\mathcal T}|_{\mathcal T})
=\{\tfrac12^{[3]},\tfrac14^{[2]},\tfrac18^{[3]}\}.
\end{equation}
The superscripts in square brackets denote multiplicities. Indeed,
the $J_\alpha$ directions have weight $1/2$, the two diagonal
quadrupole directions orthogonal to $h_0$ have weight $1/4$, and
the three cross moments have weight $1/8$. Consequently,
\begin{equation}
\operatorname{tr}_{\rm op}\sqrt{\mathsf S_{\mathcal T}}
=1+\frac9{2\sqrt2}.
\end{equation}

For an arbitrary orthonormal measurement basis $\{|b_{ix}\rangle\}_{x=1}^4$,
put $\Pi_{ix}=|b_{ix}\rangle\langle b_{ix}|$ and
$t_{ix}=\operatorname{Tr}(P_t\Pi_{ix})$. Then
$0\le t_{ix}\le1$ and $\sum_xt_{ix}=3$. Its projection onto
$\mathcal T$ is
\begin{equation}
Q_{ix}=P_t\Pi_{ix}P_t-\tfrac{t_{ix}}3P_t,
\qquad\operatorname{Tr}Q_{ix}^2=\tfrac23t_{ix}^2.
\end{equation}
The last identity uses the fact that $P_t\Pi_{ix}P_t$ has rank at
most one. The diagonal-space projector is
$4\sum_x|\Pi_{ix})(\Pi_{ix}|$; removing the identity direction
does not affect its compression to $\mathcal T$. Hence
\begin{equation}
\begin{aligned}
\operatorname{tr}_{\rm op}
(P_{\mathcal T}P_{\mathcal A_i}P_{\mathcal T})
&=\sum_x\operatorname{Tr}Q_{ix}^2\\
&=\tfrac23\sum_xt_{ix}^2\le2.
\end{aligned}
\end{equation}
For $F_{\mathcal T}=P_{\mathcal T}FP_{\mathcal T}|_{\mathcal T}$
and $f_0=(h_0,Fh_0)_0$, it follows that
\begin{equation}
\operatorname{tr}_{\rm op}F_{\mathcal T}\le2,
\qquad 0<f_0\le1.
\label{eq:collective-triplet-budget}
\end{equation}
No step assumes that the measurement basis preserves either sector.

Exact coverage makes $F_{\mathcal T}$ positive definite. For
$x\in\operatorname{ran}F$, the variational identity
\begin{equation}
(x,F^+x)_0=\sup_y\{2(x,y)_0-(y,Fy)_0\}
\end{equation}
holds even if $F$ is singular. Restricting $y$ to $\mathbb Rh_0$
or to $\mathcal T$ gives, respectively,
\begin{equation}
\begin{aligned}
(h_0,F^+h_0)_0&\ge f_0^{-1},\\
(x,F^+x)_0&\ge(x,F_{\mathcal T}^{-1}x)_0,
\quad x\in\mathcal T.
\end{aligned}
\end{equation}
Thus we need not assume that $F$ has no cross blocks between these
subspaces. Hilbert--Schmidt Cauchy's inequality further gives
\begin{equation}
\bigl(\operatorname{tr}_{\rm op}\sqrt{\mathsf S_{\mathcal T}}\bigr)^2
\le\operatorname{tr}_{\rm op}F_{\mathcal T}\,
\operatorname{tr}_{\rm op}
(\mathsf S_{\mathcal T}F_{\mathcal T}^{-1}).
\end{equation}
Combining these inequalities yields
\begin{equation}
\begin{aligned}
\Gamma
&\ge\frac4{45}\left[\frac1{4f_0}
+\operatorname{tr}_{\rm op}
(\mathsf S_{\mathcal T}F_{\mathcal T}^{-1})\right]\\
&\ge\frac4{45}\left[\frac14+
\frac12\left(1+\frac9{2\sqrt2}\right)^2\right]\\
&=\frac{31}{60}+\frac{\sqrt2}{5}.
\end{aligned}
\end{equation}
A degenerate projective measurement can be refined to a rank-one
measurement on the same system, without increasing the setting count
or losing its original estimator. The bound therefore applies to
degenerate projective measurements as well.\hfill$\square$

The bound is independent of $K$. For these nine targets with their
natural normalization and the stated Haar input-state average,
comparing it with the numerically attained four-setting value in
Table~\ref{tab:collective-comparison} gives a relative gap of
approximately $0.0341\%$. We do not
assert that the analytic bound is attained, or give a rigorous interval
certificate for the floating-point upper value. The result does not
address arbitrary POVMs, adaptive protocols, collective measurements
on several copies, or nonlinear estimators.

%% file: approximate_error.tex
Fix the measurement unitaries, estimator coefficients, and number of state copies assigned to each setting. The estimator function satisfies $\mathbb E_\rho[f_{ij}(x)]=\operatorname{Tr}(\rho O_j^{(i)})$, so
\begin{equation}
\mathbb E_\rho[\widehat o_j]
=\operatorname{Tr}(\rho\widetilde O_j)
=\operatorname{Tr}(\rho O_j)-\operatorname{Tr}(\rho R_j).
\end{equation}
We expand the squared estimation error around this mean to obtain
\begin{equation}
\begin{aligned}
\operatorname{MSE}_\rho(\widehat o_j)
&=\mathbb E_\rho\!\left[
(\widehat o_j-\operatorname{Tr}(\rho O_j))^2\right]\\
&=\operatorname{Var}_\rho(\widehat o_j)
  +[\operatorname{Tr}(\rho R_j)]^2.
\end{aligned}
\end{equation}
Independent copies give the variance in Eq.~\eqref{eq:state-variance} even for approximate reconstruction.

For Haar-distributed pure states,
\begin{equation}
\mathbb E_\psi[\rho_\psi\otimes\rho_\psi]
=\frac{I+S_{\rm sw}}{d(d+1)},
\end{equation}
where $S_{\rm sw}$ swaps the two tensor factors. Since $R_j$ is traceless and Hermitian,
\begin{equation}
\mathbb E_\psi\!\left[
  \bigl(\operatorname{Tr}(\rho_\psi R_j)\bigr)^2\right]
=\frac{\|R_j\|_F^2}{d(d+1)}.
\end{equation}
We combine this identity with the average variance $\Gamma(U,C,p)/N$ and average over targets to obtain Eq.~\eqref{eq:approximate-mse}. We evaluate $\Gamma$ with the estimator coefficients and sample allocation used in the measurement.

%% file: main.bbl
\begin{thebibliography}{53}%
\makeatletter
\providecommand \@ifxundefined [1]{%
 \@ifx{#1\undefined}
}%
\providecommand \@ifnum [1]{%
 \ifnum #1\expandafter \@firstoftwo
 \else \expandafter \@secondoftwo
 \fi
}%
\providecommand \@ifx [1]{%
 \ifx #1\expandafter \@firstoftwo
 \else \expandafter \@secondoftwo
 \fi
}%
\providecommand \natexlab [1]{#1}%
\providecommand \enquote  [1]{``#1''}%
\providecommand \bibnamefont  [1]{#1}%
\providecommand \bibfnamefont [1]{#1}%
\providecommand \citenamefont [1]{#1}%
\providecommand \href@noop [0]{\@secondoftwo}%
\providecommand \href [0]{\begingroup \@sanitize@url \@href}%
\providecommand \@href[1]{\@@startlink{#1}\@@href}%
\providecommand \@@href[1]{\endgroup#1\@@endlink}%
\providecommand \@sanitize@url [0]{\catcode `\\12\catcode `\$12\catcode
  `\&12\catcode `\#12\catcode `\^12\catcode `\_12\catcode `\%12\relax}%
\providecommand \@@startlink[1]{}%
\providecommand \@@endlink[0]{}%
\providecommand \url  [0]{\begingroup\@sanitize@url \@url }%
\providecommand \@url [1]{\endgroup\@href {#1}{\urlprefix }}%
\providecommand \urlprefix  [0]{URL }%
\providecommand \Eprint [0]{\href }%
\providecommand \doibase [0]{https://doi.org/}%
\providecommand \selectlanguage [0]{\@gobble}%
\providecommand \bibinfo  [0]{\@secondoftwo}%
\providecommand \bibfield  [0]{\@secondoftwo}%
\providecommand \translation [1]{[#1]}%
\providecommand \BibitemOpen [0]{}%
\providecommand \bibitemStop [0]{}%
\providecommand \bibitemNoStop [0]{.\EOS\space}%
\providecommand \EOS [0]{\spacefactor3000\relax}%
\providecommand \BibitemShut  [1]{\csname bibitem#1\endcsname}%
\let\auto@bib@innerbib\@empty
\bibitem [{\citenamefont {Verteletskyi}\ \emph {et~al.}(2020)\citenamefont
  {Verteletskyi}, \citenamefont {Yen},\ and\ \citenamefont
  {Izmaylov}}]{Verteletskyi2020Clique}%
  \BibitemOpen
  \bibfield  {author} {\bibinfo {author} {\bibfnamefont {V.}~\bibnamefont
  {Verteletskyi}}, \bibinfo {author} {\bibfnamefont {T.-C.}\ \bibnamefont
  {Yen}},\ and\ \bibinfo {author} {\bibfnamefont {A.~F.}\ \bibnamefont
  {Izmaylov}},\ }\bibfield  {title} {\bibinfo {title} {Measurement optimization
  in the variational quantum eigensolver using a minimum clique cover},\ }\href
  {https://doi.org/10.1063/1.5141458} {\bibfield  {journal} {\bibinfo
  {journal} {J. Chem. Phys.}\ }\textbf {\bibinfo {volume} {152}},\ \bibinfo
  {pages} {124114} (\bibinfo {year} {2020})}\BibitemShut {NoStop}%
\bibitem [{\citenamefont {Yen}\ \emph {et~al.}(2020)\citenamefont {Yen},
  \citenamefont {Verteletskyi},\ and\ \citenamefont
  {Izmaylov}}]{Yen2020Compatible}%
  \BibitemOpen
  \bibfield  {author} {\bibinfo {author} {\bibfnamefont {T.-C.}\ \bibnamefont
  {Yen}}, \bibinfo {author} {\bibfnamefont {V.}~\bibnamefont {Verteletskyi}},\
  and\ \bibinfo {author} {\bibfnamefont {A.~F.}\ \bibnamefont {Izmaylov}},\
  }\bibfield  {title} {\bibinfo {title} {Measuring all compatible operators in
  one series of single-qubit measurements using unitary transformations},\
  }\href {https://doi.org/10.1021/acs.jctc.0c00008} {\bibfield  {journal}
  {\bibinfo  {journal} {J. Chem. Theory Comput.}\ }\textbf {\bibinfo {volume}
  {16}},\ \bibinfo {pages} {2400} (\bibinfo {year} {2020})}\BibitemShut
  {NoStop}%
\bibitem [{\citenamefont {Gokhale}\ \emph {et~al.}(2020)\citenamefont
  {Gokhale}, \citenamefont {Angiuli}, \citenamefont {Ding}, \citenamefont
  {Gui}, \citenamefont {Tomesh}, \citenamefont {Suchara}, \citenamefont
  {Martonosi},\ and\ \citenamefont {Chong}}]{Gokhale2020MeasurementCost}%
  \BibitemOpen
  \bibfield  {author} {\bibinfo {author} {\bibfnamefont {P.}~\bibnamefont
  {Gokhale}}, \bibinfo {author} {\bibfnamefont {O.}~\bibnamefont {Angiuli}},
  \bibinfo {author} {\bibfnamefont {Y.}~\bibnamefont {Ding}}, \bibinfo {author}
  {\bibfnamefont {K.}~\bibnamefont {Gui}}, \bibinfo {author} {\bibfnamefont
  {T.}~\bibnamefont {Tomesh}}, \bibinfo {author} {\bibfnamefont
  {M.}~\bibnamefont {Suchara}}, \bibinfo {author} {\bibfnamefont
  {M.}~\bibnamefont {Martonosi}},\ and\ \bibinfo {author} {\bibfnamefont
  {F.~T.}\ \bibnamefont {Chong}},\ }\bibfield  {title} {\bibinfo {title}
  {{$O(N^3)$} measurement cost for variational quantum eigensolver on molecular
  {Hamiltonians}},\ }\href {https://doi.org/10.1109/TQE.2020.3035814}
  {\bibfield  {journal} {\bibinfo  {journal} {IEEE Trans. Quantum Eng.}\
  }\textbf {\bibinfo {volume} {1}},\ \bibinfo {pages} {3102324} (\bibinfo
  {year} {2020})}\BibitemShut {NoStop}%
\bibitem [{\citenamefont {Hamamura}\ and\ \citenamefont
  {Imamichi}(2020)}]{Hamamura2020Entangled}%
  \BibitemOpen
  \bibfield  {author} {\bibinfo {author} {\bibfnamefont {I.}~\bibnamefont
  {Hamamura}}\ and\ \bibinfo {author} {\bibfnamefont {T.}~\bibnamefont
  {Imamichi}},\ }\bibfield  {title} {\bibinfo {title} {Efficient evaluation of
  quantum observables using entangled measurements},\ }\href
  {https://doi.org/10.1038/s41534-020-0284-2} {\bibfield  {journal} {\bibinfo
  {journal} {npj Quantum Inf.}\ }\textbf {\bibinfo {volume} {6}},\ \bibinfo
  {pages} {56} (\bibinfo {year} {2020})}\BibitemShut {NoStop}%
\bibitem [{\citenamefont {Izmaylov}\ \emph {et~al.}(2020)\citenamefont
  {Izmaylov}, \citenamefont {Yen}, \citenamefont {Lang},\ and\ \citenamefont
  {Verteletskyi}}]{Izmaylov2020Unitary}%
  \BibitemOpen
  \bibfield  {author} {\bibinfo {author} {\bibfnamefont {A.~F.}\ \bibnamefont
  {Izmaylov}}, \bibinfo {author} {\bibfnamefont {T.-C.}\ \bibnamefont {Yen}},
  \bibinfo {author} {\bibfnamefont {R.~A.}\ \bibnamefont {Lang}},\ and\
  \bibinfo {author} {\bibfnamefont {V.}~\bibnamefont {Verteletskyi}},\
  }\bibfield  {title} {\bibinfo {title} {Unitary partitioning approach to the
  measurement problem in the variational quantum eigensolver method},\ }\href
  {https://doi.org/10.1021/acs.jctc.9b00791} {\bibfield  {journal} {\bibinfo
  {journal} {J. Chem. Theory Comput.}\ }\textbf {\bibinfo {volume} {16}},\
  \bibinfo {pages} {190} (\bibinfo {year} {2020})}\BibitemShut {NoStop}%
\bibitem [{\citenamefont {Zhao}\ \emph {et~al.}(2020)\citenamefont {Zhao},
  \citenamefont {Tranter}, \citenamefont {Kirby}, \citenamefont {Ung},
  \citenamefont {Miyake},\ and\ \citenamefont {Love}}]{Zhao2020Reduction}%
  \BibitemOpen
  \bibfield  {author} {\bibinfo {author} {\bibfnamefont {A.}~\bibnamefont
  {Zhao}}, \bibinfo {author} {\bibfnamefont {A.}~\bibnamefont {Tranter}},
  \bibinfo {author} {\bibfnamefont {W.~M.}\ \bibnamefont {Kirby}}, \bibinfo
  {author} {\bibfnamefont {S.~F.}\ \bibnamefont {Ung}}, \bibinfo {author}
  {\bibfnamefont {A.}~\bibnamefont {Miyake}},\ and\ \bibinfo {author}
  {\bibfnamefont {P.~J.}\ \bibnamefont {Love}},\ }\bibfield  {title} {\bibinfo
  {title} {Measurement reduction in variational quantum algorithms},\ }\href
  {https://doi.org/10.1103/PhysRevA.101.062322} {\bibfield  {journal} {\bibinfo
   {journal} {Phys. Rev. A}\ }\textbf {\bibinfo {volume} {101}},\ \bibinfo
  {pages} {062322} (\bibinfo {year} {2020})}\BibitemShut {NoStop}%
\bibitem [{\citenamefont {Yen}\ and\ \citenamefont
  {Izmaylov}(2021)}]{Yen2021Cartan}%
  \BibitemOpen
  \bibfield  {author} {\bibinfo {author} {\bibfnamefont {T.-C.}\ \bibnamefont
  {Yen}}\ and\ \bibinfo {author} {\bibfnamefont {A.~F.}\ \bibnamefont
  {Izmaylov}},\ }\bibfield  {title} {\bibinfo {title} {Cartan subalgebra
  approach to efficient measurements of quantum observables},\ }\href
  {https://doi.org/10.1103/PRXQuantum.2.040320} {\bibfield  {journal} {\bibinfo
   {journal} {PRX Quantum}\ }\textbf {\bibinfo {volume} {2}},\ \bibinfo {pages}
  {040320} (\bibinfo {year} {2021})}\BibitemShut {NoStop}%
\bibitem [{\citenamefont {Huggins}\ \emph {et~al.}(2021)\citenamefont
  {Huggins}, \citenamefont {McClean}, \citenamefont {Rubin}, \citenamefont
  {Jiang}, \citenamefont {Wiebe}, \citenamefont {Whaley},\ and\ \citenamefont
  {Babbush}}]{Huggins2021BasisRotation}%
  \BibitemOpen
  \bibfield  {author} {\bibinfo {author} {\bibfnamefont {W.~J.}\ \bibnamefont
  {Huggins}}, \bibinfo {author} {\bibfnamefont {J.~R.}\ \bibnamefont
  {McClean}}, \bibinfo {author} {\bibfnamefont {N.~C.}\ \bibnamefont {Rubin}},
  \bibinfo {author} {\bibfnamefont {Z.}~\bibnamefont {Jiang}}, \bibinfo
  {author} {\bibfnamefont {N.}~\bibnamefont {Wiebe}}, \bibinfo {author}
  {\bibfnamefont {K.~B.}\ \bibnamefont {Whaley}},\ and\ \bibinfo {author}
  {\bibfnamefont {R.}~\bibnamefont {Babbush}},\ }\bibfield  {title} {\bibinfo
  {title} {Efficient and noise resilient measurements for quantum chemistry on
  near-term quantum computers},\ }\href
  {https://doi.org/10.1038/s41534-020-00341-7} {\bibfield  {journal} {\bibinfo
  {journal} {npj Quantum Inf.}\ }\textbf {\bibinfo {volume} {7}},\ \bibinfo
  {pages} {23} (\bibinfo {year} {2021})}\BibitemShut {NoStop}%
\bibitem [{\citenamefont {Crawford}\ \emph {et~al.}(2021)\citenamefont
  {Crawford}, \citenamefont {van Straaten}, \citenamefont {Wang}, \citenamefont
  {Parks}, \citenamefont {Campbell},\ and\ \citenamefont
  {Brierley}}]{Crawford2021Sampling}%
  \BibitemOpen
  \bibfield  {author} {\bibinfo {author} {\bibfnamefont {O.}~\bibnamefont
  {Crawford}}, \bibinfo {author} {\bibfnamefont {B.}~\bibnamefont {van
  Straaten}}, \bibinfo {author} {\bibfnamefont {D.}~\bibnamefont {Wang}},
  \bibinfo {author} {\bibfnamefont {T.}~\bibnamefont {Parks}}, \bibinfo
  {author} {\bibfnamefont {E.}~\bibnamefont {Campbell}},\ and\ \bibinfo
  {author} {\bibfnamefont {S.}~\bibnamefont {Brierley}},\ }\bibfield  {title}
  {\bibinfo {title} {Efficient quantum measurement of {Pauli} operators in the
  presence of finite sampling error},\ }\href
  {https://doi.org/10.22331/q-2021-01-20-385} {\bibfield  {journal} {\bibinfo
  {journal} {Quantum}\ }\textbf {\bibinfo {volume} {5}},\ \bibinfo {pages}
  {385} (\bibinfo {year} {2021})}\BibitemShut {NoStop}%
\bibitem [{\citenamefont {Yen}\ \emph {et~al.}(2023)\citenamefont {Yen},
  \citenamefont {Ganeshram},\ and\ \citenamefont
  {Izmaylov}}]{Yen2023Deterministic}%
  \BibitemOpen
  \bibfield  {author} {\bibinfo {author} {\bibfnamefont {T.-C.}\ \bibnamefont
  {Yen}}, \bibinfo {author} {\bibfnamefont {A.}~\bibnamefont {Ganeshram}},\
  and\ \bibinfo {author} {\bibfnamefont {A.~F.}\ \bibnamefont {Izmaylov}},\
  }\bibfield  {title} {\bibinfo {title} {Deterministic improvements of quantum
  measurements with grouping of compatible operators, non-local
  transformations, and covariance estimates},\ }\href
  {https://doi.org/10.1038/s41534-023-00683-y} {\bibfield  {journal} {\bibinfo
  {journal} {npj Quantum Inf.}\ }\textbf {\bibinfo {volume} {9}},\ \bibinfo
  {pages} {14} (\bibinfo {year} {2023})}\BibitemShut {NoStop}%
\bibitem [{\citenamefont {Wu}\ \emph {et~al.}(2023)\citenamefont {Wu},
  \citenamefont {Sun}, \citenamefont {Huang},\ and\ \citenamefont
  {Yuan}}]{Wu2023Overlapped}%
  \BibitemOpen
  \bibfield  {author} {\bibinfo {author} {\bibfnamefont {B.}~\bibnamefont
  {Wu}}, \bibinfo {author} {\bibfnamefont {J.}~\bibnamefont {Sun}}, \bibinfo
  {author} {\bibfnamefont {Q.}~\bibnamefont {Huang}},\ and\ \bibinfo {author}
  {\bibfnamefont {X.}~\bibnamefont {Yuan}},\ }\bibfield  {title} {\bibinfo
  {title} {Overlapped grouping measurement: A unified framework for measuring
  quantum states},\ }\href {https://doi.org/10.22331/q-2023-01-13-896}
  {\bibfield  {journal} {\bibinfo  {journal} {Quantum}\ }\textbf {\bibinfo
  {volume} {7}},\ \bibinfo {pages} {896} (\bibinfo {year} {2023})}\BibitemShut
  {NoStop}%
\bibitem [{\citenamefont {Shlosberg}\ \emph {et~al.}(2023)\citenamefont
  {Shlosberg}, \citenamefont {Jena}, \citenamefont {Mukhopadhyay},
  \citenamefont {Haase}, \citenamefont {Leditzky},\ and\ \citenamefont
  {Dellantonio}}]{Shlosberg2023Adaptive}%
  \BibitemOpen
  \bibfield  {author} {\bibinfo {author} {\bibfnamefont {A.}~\bibnamefont
  {Shlosberg}}, \bibinfo {author} {\bibfnamefont {A.~J.}\ \bibnamefont {Jena}},
  \bibinfo {author} {\bibfnamefont {P.}~\bibnamefont {Mukhopadhyay}}, \bibinfo
  {author} {\bibfnamefont {J.~F.}\ \bibnamefont {Haase}}, \bibinfo {author}
  {\bibfnamefont {F.}~\bibnamefont {Leditzky}},\ and\ \bibinfo {author}
  {\bibfnamefont {L.}~\bibnamefont {Dellantonio}},\ }\bibfield  {title}
  {\bibinfo {title} {Adaptive estimation of quantum observables},\ }\href
  {https://doi.org/10.22331/q-2023-01-26-906} {\bibfield  {journal} {\bibinfo
  {journal} {Quantum}\ }\textbf {\bibinfo {volume} {7}},\ \bibinfo {pages}
  {906} (\bibinfo {year} {2023})}\BibitemShut {NoStop}%
\bibitem [{\citenamefont {Zhu}\ \emph {et~al.}(2024)\citenamefont {Zhu},
  \citenamefont {Liang}, \citenamefont {Yang},\ and\ \citenamefont
  {Li}}]{Zhu2024ShotAssignment}%
  \BibitemOpen
  \bibfield  {author} {\bibinfo {author} {\bibfnamefont {L.}~\bibnamefont
  {Zhu}}, \bibinfo {author} {\bibfnamefont {S.}~\bibnamefont {Liang}}, \bibinfo
  {author} {\bibfnamefont {C.}~\bibnamefont {Yang}},\ and\ \bibinfo {author}
  {\bibfnamefont {X.}~\bibnamefont {Li}},\ }\bibfield  {title} {\bibinfo
  {title} {Optimizing shot assignment in variational quantum eigensolver
  measurement},\ }\href {https://doi.org/10.1021/acs.jctc.3c01113} {\bibfield
  {journal} {\bibinfo  {journal} {J. Chem. Theory Comput.}\ }\textbf {\bibinfo
  {volume} {20}},\ \bibinfo {pages} {2390} (\bibinfo {year}
  {2024})}\BibitemShut {NoStop}%
\bibitem [{\citenamefont {Cotler}\ and\ \citenamefont
  {Wilczek}(2020)}]{Cotler2020Overlapping}%
  \BibitemOpen
  \bibfield  {author} {\bibinfo {author} {\bibfnamefont {J.}~\bibnamefont
  {Cotler}}\ and\ \bibinfo {author} {\bibfnamefont {F.}~\bibnamefont
  {Wilczek}},\ }\bibfield  {title} {\bibinfo {title} {Quantum overlapping
  tomography},\ }\href {https://doi.org/10.1103/PhysRevLett.124.100401}
  {\bibfield  {journal} {\bibinfo  {journal} {Phys. Rev. Lett.}\ }\textbf
  {\bibinfo {volume} {124}},\ \bibinfo {pages} {100401} (\bibinfo {year}
  {2020})}\BibitemShut {NoStop}%
\bibitem [{\citenamefont {Bonet-Monroig}\ \emph {et~al.}(2020)\citenamefont
  {Bonet-Monroig}, \citenamefont {Babbush},\ and\ \citenamefont
  {O'Brien}}]{BonetMonroig2020PartialTomography}%
  \BibitemOpen
  \bibfield  {author} {\bibinfo {author} {\bibfnamefont {X.}~\bibnamefont
  {Bonet-Monroig}}, \bibinfo {author} {\bibfnamefont {R.}~\bibnamefont
  {Babbush}},\ and\ \bibinfo {author} {\bibfnamefont {T.~E.}\ \bibnamefont
  {O'Brien}},\ }\bibfield  {title} {\bibinfo {title} {Nearly optimal
  measurement scheduling for partial tomography of quantum states},\ }\href
  {https://doi.org/10.1103/PhysRevX.10.031064} {\bibfield  {journal} {\bibinfo
  {journal} {Phys. Rev. X}\ }\textbf {\bibinfo {volume} {10}},\ \bibinfo
  {pages} {031064} (\bibinfo {year} {2020})}\BibitemShut {NoStop}%
\bibitem [{\citenamefont {Aaronson}(2018)}]{Aaronson2018ShadowTomography}%
  \BibitemOpen
  \bibfield  {author} {\bibinfo {author} {\bibfnamefont {S.}~\bibnamefont
  {Aaronson}},\ }\bibfield  {title} {\bibinfo {title} {Shadow tomography of
  quantum states},\ }in\ \href {https://doi.org/10.1145/3188745.3188802} {\emph
  {\bibinfo {booktitle} {Proceedings of the 50th Annual ACM SIGACT Symposium on
  Theory of Computing}}},\ \bibinfo {series and number} {STOC 2018}\ (\bibinfo
  {publisher} {Association for Computing Machinery},\ \bibinfo {address} {New
  York},\ \bibinfo {year} {2018})\ pp.\ \bibinfo {pages} {325--338}\BibitemShut
  {NoStop}%
\bibitem [{\citenamefont {Huang}\ \emph {et~al.}(2020)\citenamefont {Huang},
  \citenamefont {Kueng},\ and\ \citenamefont {Preskill}}]{Huang2020Shadows}%
  \BibitemOpen
  \bibfield  {author} {\bibinfo {author} {\bibfnamefont {H.-Y.}\ \bibnamefont
  {Huang}}, \bibinfo {author} {\bibfnamefont {R.}~\bibnamefont {Kueng}},\ and\
  \bibinfo {author} {\bibfnamefont {J.}~\bibnamefont {Preskill}},\ }\bibfield
  {title} {\bibinfo {title} {Predicting many properties of a quantum system
  from very few measurements},\ }\href
  {https://doi.org/10.1038/s41567-020-0932-7} {\bibfield  {journal} {\bibinfo
  {journal} {Nat. Phys.}\ }\textbf {\bibinfo {volume} {16}},\ \bibinfo {pages}
  {1050} (\bibinfo {year} {2020})}\BibitemShut {NoStop}%
\bibitem [{\citenamefont {Elben}\ \emph {et~al.}(2023)\citenamefont {Elben},
  \citenamefont {Flammia}, \citenamefont {Huang}, \citenamefont {Kueng},
  \citenamefont {Preskill}, \citenamefont {Vermersch},\ and\ \citenamefont
  {Zoller}}]{Elben2023Toolbox}%
  \BibitemOpen
  \bibfield  {author} {\bibinfo {author} {\bibfnamefont {A.}~\bibnamefont
  {Elben}}, \bibinfo {author} {\bibfnamefont {S.~T.}\ \bibnamefont {Flammia}},
  \bibinfo {author} {\bibfnamefont {H.-Y.}\ \bibnamefont {Huang}}, \bibinfo
  {author} {\bibfnamefont {R.}~\bibnamefont {Kueng}}, \bibinfo {author}
  {\bibfnamefont {J.}~\bibnamefont {Preskill}}, \bibinfo {author}
  {\bibfnamefont {B.}~\bibnamefont {Vermersch}},\ and\ \bibinfo {author}
  {\bibfnamefont {P.}~\bibnamefont {Zoller}},\ }\bibfield  {title} {\bibinfo
  {title} {The randomized measurement toolbox},\ }\href
  {https://doi.org/10.1038/s42254-022-00535-2} {\bibfield  {journal} {\bibinfo
  {journal} {Nat. Rev. Phys.}\ }\textbf {\bibinfo {volume} {5}},\ \bibinfo
  {pages} {9} (\bibinfo {year} {2023})}\BibitemShut {NoStop}%
\bibitem [{\citenamefont {Hadfield}\ \emph {et~al.}(2022)\citenamefont
  {Hadfield}, \citenamefont {Bravyi}, \citenamefont {Raymond},\ and\
  \citenamefont {Mezzacapo}}]{Hadfield2022LocallyBiased}%
  \BibitemOpen
  \bibfield  {author} {\bibinfo {author} {\bibfnamefont {C.}~\bibnamefont
  {Hadfield}}, \bibinfo {author} {\bibfnamefont {S.}~\bibnamefont {Bravyi}},
  \bibinfo {author} {\bibfnamefont {R.}~\bibnamefont {Raymond}},\ and\ \bibinfo
  {author} {\bibfnamefont {A.}~\bibnamefont {Mezzacapo}},\ }\bibfield  {title}
  {\bibinfo {title} {Measurements of quantum hamiltonians with locally-biased
  classical shadows},\ }\href {https://doi.org/10.1007/s00220-022-04343-8}
  {\bibfield  {journal} {\bibinfo  {journal} {Commun. Math. Phys.}\ }\textbf
  {\bibinfo {volume} {391}},\ \bibinfo {pages} {951} (\bibinfo {year}
  {2022})}\BibitemShut {NoStop}%
\bibitem [{\citenamefont {Huang}\ \emph {et~al.}(2021)\citenamefont {Huang},
  \citenamefont {Kueng},\ and\ \citenamefont
  {Preskill}}]{Huang2021Derandomization}%
  \BibitemOpen
  \bibfield  {author} {\bibinfo {author} {\bibfnamefont {H.-Y.}\ \bibnamefont
  {Huang}}, \bibinfo {author} {\bibfnamefont {R.}~\bibnamefont {Kueng}},\ and\
  \bibinfo {author} {\bibfnamefont {J.}~\bibnamefont {Preskill}},\ }\bibfield
  {title} {\bibinfo {title} {Efficient estimation of {Pauli} observables by
  derandomization},\ }\href {https://doi.org/10.1103/PhysRevLett.127.030503}
  {\bibfield  {journal} {\bibinfo  {journal} {Phys. Rev. Lett.}\ }\textbf
  {\bibinfo {volume} {127}},\ \bibinfo {pages} {030503} (\bibinfo {year}
  {2021})}\BibitemShut {NoStop}%
\bibitem [{\citenamefont {Garc{\'i}a-P{\'e}rez}\ \emph
  {et~al.}(2021)\citenamefont {Garc{\'i}a-P{\'e}rez}, \citenamefont {Rossi},
  \citenamefont {Sokolov}, \citenamefont {Tacchino}, \citenamefont
  {Barkoutsos}, \citenamefont {Mazzola}, \citenamefont {Tavernelli},\ and\
  \citenamefont {Maniscalco}}]{GarciaPerez2021Learning}%
  \BibitemOpen
  \bibfield  {author} {\bibinfo {author} {\bibfnamefont {G.}~\bibnamefont
  {Garc{\'i}a-P{\'e}rez}}, \bibinfo {author} {\bibfnamefont {M.~A.~C.}\
  \bibnamefont {Rossi}}, \bibinfo {author} {\bibfnamefont {B.}~\bibnamefont
  {Sokolov}}, \bibinfo {author} {\bibfnamefont {F.}~\bibnamefont {Tacchino}},
  \bibinfo {author} {\bibfnamefont {P.~K.}\ \bibnamefont {Barkoutsos}},
  \bibinfo {author} {\bibfnamefont {G.}~\bibnamefont {Mazzola}}, \bibinfo
  {author} {\bibfnamefont {I.}~\bibnamefont {Tavernelli}},\ and\ \bibinfo
  {author} {\bibfnamefont {S.}~\bibnamefont {Maniscalco}},\ }\bibfield  {title}
  {\bibinfo {title} {Learning to measure: Adaptive informationally complete
  generalized measurements for quantum algorithms},\ }\href
  {https://doi.org/10.1103/PRXQuantum.2.040342} {\bibfield  {journal} {\bibinfo
   {journal} {PRX Quantum}\ }\textbf {\bibinfo {volume} {2}},\ \bibinfo {pages}
  {040342} (\bibinfo {year} {2021})}\BibitemShut {NoStop}%
\bibitem [{\citenamefont {Nguyen}\ \emph {et~al.}(2022)\citenamefont {Nguyen},
  \citenamefont {B{\"o}nsel}, \citenamefont {Steinberg},\ and\ \citenamefont
  {G{\"u}hne}}]{Nguyen2022Generalized}%
  \BibitemOpen
  \bibfield  {author} {\bibinfo {author} {\bibfnamefont {H.~C.}\ \bibnamefont
  {Nguyen}}, \bibinfo {author} {\bibfnamefont {J.~L.}\ \bibnamefont
  {B{\"o}nsel}}, \bibinfo {author} {\bibfnamefont {J.}~\bibnamefont
  {Steinberg}},\ and\ \bibinfo {author} {\bibfnamefont {O.}~\bibnamefont
  {G{\"u}hne}},\ }\bibfield  {title} {\bibinfo {title} {Optimizing shadow
  tomography with generalized measurements},\ }\href
  {https://doi.org/10.1103/PhysRevLett.129.220502} {\bibfield  {journal}
  {\bibinfo  {journal} {Phys. Rev. Lett.}\ }\textbf {\bibinfo {volume} {129}},\
  \bibinfo {pages} {220502} (\bibinfo {year} {2022})}\BibitemShut {NoStop}%
\bibitem [{\citenamefont {McNulty}\ \emph {et~al.}(2023)\citenamefont
  {McNulty}, \citenamefont {Maciejewski},\ and\ \citenamefont
  {Oszmaniec}}]{McNulty2023Joint}%
  \BibitemOpen
  \bibfield  {author} {\bibinfo {author} {\bibfnamefont {D.}~\bibnamefont
  {McNulty}}, \bibinfo {author} {\bibfnamefont {F.~B.}\ \bibnamefont
  {Maciejewski}},\ and\ \bibinfo {author} {\bibfnamefont {M.}~\bibnamefont
  {Oszmaniec}},\ }\bibfield  {title} {\bibinfo {title} {Estimating quantum
  hamiltonians via joint measurements of noisy noncommuting observables},\
  }\href {https://doi.org/10.1103/PhysRevLett.130.100801} {\bibfield  {journal}
  {\bibinfo  {journal} {Phys. Rev. Lett.}\ }\textbf {\bibinfo {volume} {130}},\
  \bibinfo {pages} {100801} (\bibinfo {year} {2023})}\BibitemShut {NoStop}%
\bibitem [{\citenamefont {Acharya}\ \emph {et~al.}(2021)\citenamefont
  {Acharya}, \citenamefont {Saha},\ and\ \citenamefont
  {Sengupta}}]{Acharya2021POVMShadows}%
  \BibitemOpen
  \bibfield  {author} {\bibinfo {author} {\bibfnamefont {A.}~\bibnamefont
  {Acharya}}, \bibinfo {author} {\bibfnamefont {S.}~\bibnamefont {Saha}},\ and\
  \bibinfo {author} {\bibfnamefont {A.~M.}\ \bibnamefont {Sengupta}},\
  }\bibfield  {title} {\bibinfo {title} {Shadow tomography based on
  informationally complete positive operator-valued measure},\ }\href
  {https://doi.org/10.1103/PhysRevA.104.052418} {\bibfield  {journal} {\bibinfo
   {journal} {Phys. Rev. A}\ }\textbf {\bibinfo {volume} {104}},\ \bibinfo
  {pages} {052418} (\bibinfo {year} {2021})}\BibitemShut {NoStop}%
\bibitem [{\citenamefont {Hu}\ \emph {et~al.}(2023)\citenamefont {Hu},
  \citenamefont {Choi},\ and\ \citenamefont {You}}]{Hu2023LocallyScrambled}%
  \BibitemOpen
  \bibfield  {author} {\bibinfo {author} {\bibfnamefont {H.-Y.}\ \bibnamefont
  {Hu}}, \bibinfo {author} {\bibfnamefont {S.}~\bibnamefont {Choi}},\ and\
  \bibinfo {author} {\bibfnamefont {Y.-Z.}\ \bibnamefont {You}},\ }\bibfield
  {title} {\bibinfo {title} {Classical shadow tomography with locally scrambled
  quantum dynamics},\ }\href {https://doi.org/10.1103/PhysRevResearch.5.023027}
  {\bibfield  {journal} {\bibinfo  {journal} {Phys. Rev. Research}\ }\textbf
  {\bibinfo {volume} {5}},\ \bibinfo {pages} {023027} (\bibinfo {year}
  {2023})}\BibitemShut {NoStop}%
\bibitem [{\citenamefont {Ippoliti}\ \emph {et~al.}(2023)\citenamefont
  {Ippoliti}, \citenamefont {Li}, \citenamefont {Rakovszky},\ and\
  \citenamefont {Khemani}}]{Ippoliti2023OptimalDepth}%
  \BibitemOpen
  \bibfield  {author} {\bibinfo {author} {\bibfnamefont {M.}~\bibnamefont
  {Ippoliti}}, \bibinfo {author} {\bibfnamefont {Y.}~\bibnamefont {Li}},
  \bibinfo {author} {\bibfnamefont {T.}~\bibnamefont {Rakovszky}},\ and\
  \bibinfo {author} {\bibfnamefont {V.}~\bibnamefont {Khemani}},\ }\bibfield
  {title} {\bibinfo {title} {Operator relaxation and the optimal depth of
  classical shadows},\ }\href {https://doi.org/10.1103/PhysRevLett.130.230403}
  {\bibfield  {journal} {\bibinfo  {journal} {Phys. Rev. Lett.}\ }\textbf
  {\bibinfo {volume} {130}},\ \bibinfo {pages} {230403} (\bibinfo {year}
  {2023})}\BibitemShut {NoStop}%
\bibitem [{\citenamefont {Bertoni}\ \emph {et~al.}(2024)\citenamefont
  {Bertoni}, \citenamefont {Haferkamp}, \citenamefont {Hinsche}, \citenamefont
  {Ioannou}, \citenamefont {Eisert},\ and\ \citenamefont
  {Pashayan}}]{Bertoni2024Shallow}%
  \BibitemOpen
  \bibfield  {author} {\bibinfo {author} {\bibfnamefont {C.}~\bibnamefont
  {Bertoni}}, \bibinfo {author} {\bibfnamefont {J.}~\bibnamefont {Haferkamp}},
  \bibinfo {author} {\bibfnamefont {M.}~\bibnamefont {Hinsche}}, \bibinfo
  {author} {\bibfnamefont {M.}~\bibnamefont {Ioannou}}, \bibinfo {author}
  {\bibfnamefont {J.}~\bibnamefont {Eisert}},\ and\ \bibinfo {author}
  {\bibfnamefont {H.}~\bibnamefont {Pashayan}},\ }\bibfield  {title} {\bibinfo
  {title} {Shallow shadows: Expectation estimation using low-depth random
  {Clifford} circuits},\ }\href
  {https://doi.org/10.1103/PhysRevLett.133.020602} {\bibfield  {journal}
  {\bibinfo  {journal} {Phys. Rev. Lett.}\ }\textbf {\bibinfo {volume} {133}},\
  \bibinfo {pages} {020602} (\bibinfo {year} {2024})}\BibitemShut {NoStop}%
\bibitem [{\citenamefont {Rozon}\ \emph {et~al.}(2024)\citenamefont {Rozon},
  \citenamefont {Bao},\ and\ \citenamefont {Agarwal}}]{Rozon2024Noise}%
  \BibitemOpen
  \bibfield  {author} {\bibinfo {author} {\bibfnamefont {P.-G.}\ \bibnamefont
  {Rozon}}, \bibinfo {author} {\bibfnamefont {N.}~\bibnamefont {Bao}},\ and\
  \bibinfo {author} {\bibfnamefont {K.}~\bibnamefont {Agarwal}},\ }\bibfield
  {title} {\bibinfo {title} {Optimal twirling depth for classical shadows in
  the presence of noise},\ }\href
  {https://doi.org/10.1103/PhysRevLett.133.130803} {\bibfield  {journal}
  {\bibinfo  {journal} {Phys. Rev. Lett.}\ }\textbf {\bibinfo {volume} {133}},\
  \bibinfo {pages} {130803} (\bibinfo {year} {2024})}\BibitemShut {NoStop}%
\bibitem [{\citenamefont {D'Ariano}\ and\ \citenamefont
  {Perinotti}(2007)}]{DAriano2007Optimal}%
  \BibitemOpen
  \bibfield  {author} {\bibinfo {author} {\bibfnamefont {G.~M.}\ \bibnamefont
  {D'Ariano}}\ and\ \bibinfo {author} {\bibfnamefont {P.}~\bibnamefont
  {Perinotti}},\ }\bibfield  {title} {\bibinfo {title} {Optimal data processing
  for quantum measurements},\ }\href
  {https://doi.org/10.1103/PhysRevLett.98.020403} {\bibfield  {journal}
  {\bibinfo  {journal} {Phys. Rev. Lett.}\ }\textbf {\bibinfo {volume} {98}},\
  \bibinfo {pages} {020403} (\bibinfo {year} {2007})}\BibitemShut {NoStop}%
\bibitem [{\citenamefont {Fischer}\ \emph {et~al.}(2024)\citenamefont
  {Fischer}, \citenamefont {Dao}, \citenamefont {Tavernelli},\ and\
  \citenamefont {Tacchino}}]{Fischer2024DualFrame}%
  \BibitemOpen
  \bibfield  {author} {\bibinfo {author} {\bibfnamefont {L.~E.}\ \bibnamefont
  {Fischer}}, \bibinfo {author} {\bibfnamefont {T.}~\bibnamefont {Dao}},
  \bibinfo {author} {\bibfnamefont {I.}~\bibnamefont {Tavernelli}},\ and\
  \bibinfo {author} {\bibfnamefont {F.}~\bibnamefont {Tacchino}},\ }\bibfield
  {title} {\bibinfo {title} {Dual-frame optimization for informationally
  complete quantum measurements},\ }\href
  {https://doi.org/10.1103/PhysRevA.109.062415} {\bibfield  {journal} {\bibinfo
   {journal} {Phys. Rev. A}\ }\textbf {\bibinfo {volume} {109}},\ \bibinfo
  {pages} {062415} (\bibinfo {year} {2024})}\BibitemShut {NoStop}%
\bibitem [{\citenamefont {Malmi}\ \emph {et~al.}(2024)\citenamefont {Malmi},
  \citenamefont {Korhonen}, \citenamefont {Cavalcanti},\ and\ \citenamefont
  {Garc{\'i}a-P{\'e}rez}}]{Malmi2024Enhanced}%
  \BibitemOpen
  \bibfield  {author} {\bibinfo {author} {\bibfnamefont {J.}~\bibnamefont
  {Malmi}}, \bibinfo {author} {\bibfnamefont {K.}~\bibnamefont {Korhonen}},
  \bibinfo {author} {\bibfnamefont {D.}~\bibnamefont {Cavalcanti}},\ and\
  \bibinfo {author} {\bibfnamefont {G.}~\bibnamefont {Garc{\'i}a-P{\'e}rez}},\
  }\bibfield  {title} {\bibinfo {title} {Enhanced observable estimation through
  classical optimization of informationally overcomplete measurement data:
  Beyond classical shadows},\ }\href
  {https://doi.org/10.1103/PhysRevA.109.062412} {\bibfield  {journal} {\bibinfo
   {journal} {Phys. Rev. A}\ }\textbf {\bibinfo {volume} {109}},\ \bibinfo
  {pages} {062412} (\bibinfo {year} {2024})}\BibitemShut {NoStop}%
\bibitem [{\citenamefont {Cai}\ \emph {et~al.}(2025)\citenamefont {Cai},
  \citenamefont {Chapman}, \citenamefont {Jnane},\ and\ \citenamefont
  {Koczor}}]{Cai2025Biased}%
  \BibitemOpen
  \bibfield  {author} {\bibinfo {author} {\bibfnamefont {Z.}~\bibnamefont
  {Cai}}, \bibinfo {author} {\bibfnamefont {A.}~\bibnamefont {Chapman}},
  \bibinfo {author} {\bibfnamefont {H.}~\bibnamefont {Jnane}},\ and\ \bibinfo
  {author} {\bibfnamefont {B.}~\bibnamefont {Koczor}},\ }\bibfield  {title}
  {\bibinfo {title} {Biased estimator channels for classical shadows},\ }\href
  {https://doi.org/10.1103/PhysRevA.111.L030402} {\bibfield  {journal}
  {\bibinfo  {journal} {Phys. Rev. A}\ }\textbf {\bibinfo {volume} {111}},\
  \bibinfo {pages} {L030402} (\bibinfo {year} {2025})}\BibitemShut {NoStop}%
\bibitem [{\citenamefont {Wu}\ \emph {et~al.}(2024)\citenamefont {Wu},
  \citenamefont {Yao}, \citenamefont {Zhang},\ and\ \citenamefont
  {Li}}]{Wu2024Randomness}%
  \BibitemOpen
  \bibfield  {author} {\bibinfo {author} {\bibfnamefont {Y.}~\bibnamefont
  {Wu}}, \bibinfo {author} {\bibfnamefont {J.}~\bibnamefont {Yao}}, \bibinfo
  {author} {\bibfnamefont {P.}~\bibnamefont {Zhang}},\ and\ \bibinfo {author}
  {\bibfnamefont {X.}~\bibnamefont {Li}},\ }\bibfield  {title} {\bibinfo
  {title} {Randomness-enhanced expressivity of quantum neural networks},\
  }\href {https://doi.org/10.1103/PhysRevLett.132.010602} {\bibfield  {journal}
  {\bibinfo  {journal} {Phys. Rev. Lett.}\ }\textbf {\bibinfo {volume} {132}},\
  \bibinfo {pages} {010602} (\bibinfo {year} {2024})}\BibitemShut {NoStop}%
\bibitem [{\citenamefont {Yang}\ \emph {et~al.}(2025)\citenamefont {Yang},
  \citenamefont {Huang},\ and\ \citenamefont {Zhou}}]{Yang2025Incompatible}%
  \BibitemOpen
  \bibfield  {author} {\bibinfo {author} {\bibfnamefont {M.}~\bibnamefont
  {Yang}}, \bibinfo {author} {\bibfnamefont {Y.}~\bibnamefont {Huang}},\ and\
  \bibinfo {author} {\bibfnamefont {D.~L.}\ \bibnamefont {Zhou}},\ }\bibfield
  {title} {\bibinfo {title} {Measuring incompatible observables with quantum
  neural networks},\ }\href {https://doi.org/10.1103/qz9c-m3z4} {\bibfield
  {journal} {\bibinfo  {journal} {Phys. Rev. B}\ }\textbf {\bibinfo {volume}
  {112}},\ \bibinfo {pages} {214308} (\bibinfo {year} {2025})}\BibitemShut
  {NoStop}%
\bibitem [{\citenamefont {Yang}\ \emph {et~al.}(2026)\citenamefont {Yang},
  \citenamefont {Chen},\ and\ \citenamefont {Zhu}}]{Yang2026MinimalSettings}%
  \BibitemOpen
  \bibfield  {author} {\bibinfo {author} {\bibfnamefont {Z.}~\bibnamefont
  {Yang}}, \bibinfo {author} {\bibfnamefont {D.}~\bibnamefont {Chen}},\ and\
  \bibinfo {author} {\bibfnamefont {H.}~\bibnamefont {Zhu}},\ }\bibfield
  {title} {\bibinfo {title} {Optimal shadow estimation with minimal measurement
  settings},\ }\href {https://doi.org/10.1103/g4lh-56r1} {\bibfield  {journal}
  {\bibinfo  {journal} {Phys. Rev. Lett.}\ }\textbf {\bibinfo {volume} {137}},\
  \bibinfo {pages} {140804} (\bibinfo {year} {2026})}\BibitemShut {NoStop}%
\bibitem [{\citenamefont {Golub}\ and\ \citenamefont
  {Pereyra}(1973)}]{Golub1973VariableProjection}%
  \BibitemOpen
  \bibfield  {author} {\bibinfo {author} {\bibfnamefont {G.~H.}\ \bibnamefont
  {Golub}}\ and\ \bibinfo {author} {\bibfnamefont {V.}~\bibnamefont
  {Pereyra}},\ }\bibfield  {title} {\bibinfo {title} {The differentiation of
  pseudo-inverses and nonlinear least squares problems whose variables
  separate},\ }\href {https://doi.org/10.1137/0710036} {\bibfield  {journal}
  {\bibinfo  {journal} {SIAM J. Numer. Anal.}\ }\textbf {\bibinfo {volume}
  {10}},\ \bibinfo {pages} {413} (\bibinfo {year} {1973})}\BibitemShut
  {NoStop}%
\bibitem [{\citenamefont {Wootters}\ and\ \citenamefont
  {Fields}(1989)}]{Wootters1989MUB}%
  \BibitemOpen
  \bibfield  {author} {\bibinfo {author} {\bibfnamefont {W.~K.}\ \bibnamefont
  {Wootters}}\ and\ \bibinfo {author} {\bibfnamefont {B.~D.}\ \bibnamefont
  {Fields}},\ }\bibfield  {title} {\bibinfo {title} {Optimal
  state-determination by mutually unbiased measurements},\ }\href
  {https://doi.org/10.1016/0003-4916(89)90322-9} {\bibfield  {journal}
  {\bibinfo  {journal} {Ann. Phys.}\ }\textbf {\bibinfo {volume} {191}},\
  \bibinfo {pages} {363} (\bibinfo {year} {1989})}\BibitemShut {NoStop}%
\bibitem [{\citenamefont {T{\'o}th}\ \emph {et~al.}(2010)\citenamefont
  {T{\'o}th}, \citenamefont {Wieczorek}, \citenamefont {Gross}, \citenamefont
  {Krischek}, \citenamefont {Schwemmer},\ and\ \citenamefont
  {Weinfurter}}]{Toth2010}%
  \BibitemOpen
  \bibfield  {author} {\bibinfo {author} {\bibfnamefont {G.}~\bibnamefont
  {T{\'o}th}}, \bibinfo {author} {\bibfnamefont {W.}~\bibnamefont {Wieczorek}},
  \bibinfo {author} {\bibfnamefont {D.}~\bibnamefont {Gross}}, \bibinfo
  {author} {\bibfnamefont {R.}~\bibnamefont {Krischek}}, \bibinfo {author}
  {\bibfnamefont {C.}~\bibnamefont {Schwemmer}},\ and\ \bibinfo {author}
  {\bibfnamefont {H.}~\bibnamefont {Weinfurter}},\ }\bibfield  {title}
  {\bibinfo {title} {Permutationally invariant quantum tomography},\ }\href
  {https://doi.org/10.1103/PhysRevLett.105.250403} {\bibfield  {journal}
  {\bibinfo  {journal} {Phys. Rev. Lett.}\ }\textbf {\bibinfo {volume} {105}},\
  \bibinfo {pages} {250403} (\bibinfo {year} {2010})}\BibitemShut {NoStop}%
\bibitem [{\citenamefont {Moroder}\ \emph {et~al.}(2012)\citenamefont
  {Moroder}, \citenamefont {Hyllus}, \citenamefont {T{\'o}th}, \citenamefont
  {Schwemmer}, \citenamefont {Niggebaum}, \citenamefont {Gaile}, \citenamefont
  {G{\"u}hne},\ and\ \citenamefont {Weinfurter}}]{Moroder2012}%
  \BibitemOpen
  \bibfield  {author} {\bibinfo {author} {\bibfnamefont {T.}~\bibnamefont
  {Moroder}}, \bibinfo {author} {\bibfnamefont {P.}~\bibnamefont {Hyllus}},
  \bibinfo {author} {\bibfnamefont {G.}~\bibnamefont {T{\'o}th}}, \bibinfo
  {author} {\bibfnamefont {C.}~\bibnamefont {Schwemmer}}, \bibinfo {author}
  {\bibfnamefont {A.}~\bibnamefont {Niggebaum}}, \bibinfo {author}
  {\bibfnamefont {S.}~\bibnamefont {Gaile}}, \bibinfo {author} {\bibfnamefont
  {O.}~\bibnamefont {G{\"u}hne}},\ and\ \bibinfo {author} {\bibfnamefont
  {H.}~\bibnamefont {Weinfurter}},\ }\bibfield  {title} {\bibinfo {title}
  {Permutationally invariant state reconstruction},\ }\href
  {https://doi.org/10.1088/1367-2630/14/10/105001} {\bibfield  {journal}
  {\bibinfo  {journal} {New J. Phys.}\ }\textbf {\bibinfo {volume} {14}},\
  \bibinfo {pages} {105001} (\bibinfo {year} {2012})}\BibitemShut {NoStop}%
\bibitem [{\citenamefont {Klimov}\ \emph {et~al.}(2013)\citenamefont {Klimov},
  \citenamefont {Bj{\"o}rk},\ and\ \citenamefont
  {S{\'a}nchez-Soto}}]{Klimov2013}%
  \BibitemOpen
  \bibfield  {author} {\bibinfo {author} {\bibfnamefont {A.~B.}\ \bibnamefont
  {Klimov}}, \bibinfo {author} {\bibfnamefont {G.}~\bibnamefont {Bj{\"o}rk}},\
  and\ \bibinfo {author} {\bibfnamefont {L.~L.}\ \bibnamefont
  {S{\'a}nchez-Soto}},\ }\bibfield  {title} {\bibinfo {title} {Optimal quantum
  tomography of permutationally invariant qubits},\ }\href
  {https://doi.org/10.1103/PhysRevA.87.012109} {\bibfield  {journal} {\bibinfo
  {journal} {Phys. Rev. A}\ }\textbf {\bibinfo {volume} {87}},\ \bibinfo
  {pages} {012109} (\bibinfo {year} {2013})}\BibitemShut {NoStop}%
\bibitem [{\citenamefont {Watrous}(2018)}]{Watrous2018}%
  \BibitemOpen
  \bibfield  {author} {\bibinfo {author} {\bibfnamefont {J.}~\bibnamefont
  {Watrous}},\ }\href {https://doi.org/10.1017/9781316848142} {\emph {\bibinfo
  {title} {The Theory of Quantum Information}}}\ (\bibinfo  {publisher}
  {Cambridge University Press},\ \bibinfo {address} {Cambridge},\ \bibinfo
  {year} {2018})\BibitemShut {NoStop}%
\bibitem [{\citenamefont {Zhu}(2017)}]{Zhu2017Designs}%
  \BibitemOpen
  \bibfield  {author} {\bibinfo {author} {\bibfnamefont {H.}~\bibnamefont
  {Zhu}},\ }\bibfield  {title} {\bibinfo {title} {Multiqubit {Clifford} groups
  are unitary 3-designs},\ }\href {https://doi.org/10.1103/PhysRevA.96.062336}
  {\bibfield  {journal} {\bibinfo  {journal} {Phys. Rev. A}\ }\textbf {\bibinfo
  {volume} {96}},\ \bibinfo {pages} {062336} (\bibinfo {year}
  {2017})}\BibitemShut {NoStop}%
\bibitem [{\citenamefont {Koh}\ and\ \citenamefont
  {Grewal}(2022)}]{Koh2022NoisyShadows}%
  \BibitemOpen
  \bibfield  {author} {\bibinfo {author} {\bibfnamefont {D.~E.}\ \bibnamefont
  {Koh}}\ and\ \bibinfo {author} {\bibfnamefont {S.}~\bibnamefont {Grewal}},\
  }\bibfield  {title} {\bibinfo {title} {Classical shadows with noise},\ }\href
  {https://doi.org/10.22331/q-2022-08-16-776} {\bibfield  {journal} {\bibinfo
  {journal} {Quantum}\ }\textbf {\bibinfo {volume} {6}},\ \bibinfo {pages}
  {776} (\bibinfo {year} {2022})}\BibitemShut {NoStop}%
\bibitem [{\citenamefont {Chen}\ \emph {et~al.}(2021)\citenamefont {Chen},
  \citenamefont {Yu}, \citenamefont {Zeng},\ and\ \citenamefont
  {Flammia}}]{Chen2021RobustShadows}%
  \BibitemOpen
  \bibfield  {author} {\bibinfo {author} {\bibfnamefont {S.}~\bibnamefont
  {Chen}}, \bibinfo {author} {\bibfnamefont {W.}~\bibnamefont {Yu}}, \bibinfo
  {author} {\bibfnamefont {P.}~\bibnamefont {Zeng}},\ and\ \bibinfo {author}
  {\bibfnamefont {S.~T.}\ \bibnamefont {Flammia}},\ }\bibfield  {title}
  {\bibinfo {title} {Robust shadow estimation},\ }\href
  {https://doi.org/10.1103/PRXQuantum.2.030348} {\bibfield  {journal} {\bibinfo
   {journal} {PRX Quantum}\ }\textbf {\bibinfo {volume} {2}},\ \bibinfo {pages}
  {030348} (\bibinfo {year} {2021})}\BibitemShut {NoStop}%
\bibitem [{\citenamefont {Scott}(2006)}]{Scott2006Tight}%
  \BibitemOpen
  \bibfield  {author} {\bibinfo {author} {\bibfnamefont {A.~J.}\ \bibnamefont
  {Scott}},\ }\bibfield  {title} {\bibinfo {title} {Tight informationally
  complete quantum measurements},\ }\href
  {https://doi.org/10.1088/0305-4470/39/43/009} {\bibfield  {journal} {\bibinfo
   {journal} {J. Phys. A: Math. Gen.}\ }\textbf {\bibinfo {volume} {39}},\
  \bibinfo {pages} {13507} (\bibinfo {year} {2006})}\BibitemShut {NoStop}%
\bibitem [{\citenamefont {Reggio}\ \emph {et~al.}(2024)\citenamefont {Reggio},
  \citenamefont {Butt}, \citenamefont {Lytle},\ and\ \citenamefont
  {Draper}}]{Reggio2024Partitioning}%
  \BibitemOpen
  \bibfield  {author} {\bibinfo {author} {\bibfnamefont {B.}~\bibnamefont
  {Reggio}}, \bibinfo {author} {\bibfnamefont {N.}~\bibnamefont {Butt}},
  \bibinfo {author} {\bibfnamefont {A.}~\bibnamefont {Lytle}},\ and\ \bibinfo
  {author} {\bibfnamefont {P.}~\bibnamefont {Draper}},\ }\bibfield  {title}
  {\bibinfo {title} {Fast partitioning of {Pauli} strings into commuting
  families for optimal expectation value measurements of dense operators},\
  }\href {https://doi.org/10.1103/PhysRevA.110.022606} {\bibfield  {journal}
  {\bibinfo  {journal} {Phys. Rev. A}\ }\textbf {\bibinfo {volume} {110}},\
  \bibinfo {pages} {022606} (\bibinfo {year} {2024})}\BibitemShut {NoStop}%
\bibitem [{\citenamefont {Choi}\ \emph {et~al.}(2023)\citenamefont {Choi},
  \citenamefont {Loaiza},\ and\ \citenamefont
  {Izmaylov}}]{Choi2023FluidFragments}%
  \BibitemOpen
  \bibfield  {author} {\bibinfo {author} {\bibfnamefont {S.}~\bibnamefont
  {Choi}}, \bibinfo {author} {\bibfnamefont {I.}~\bibnamefont {Loaiza}},\ and\
  \bibinfo {author} {\bibfnamefont {A.~F.}\ \bibnamefont {Izmaylov}},\
  }\bibfield  {title} {\bibinfo {title} {Fluid fermionic fragments for
  optimizing quantum measurements of electronic {Hamiltonians} in the
  variational quantum eigensolver},\ }\href
  {https://doi.org/10.22331/q-2023-01-03-889} {\bibfield  {journal} {\bibinfo
  {journal} {Quantum}\ }\textbf {\bibinfo {volume} {7}},\ \bibinfo {pages}
  {889} (\bibinfo {year} {2023})}\BibitemShut {NoStop}%
\bibitem [{\citenamefont {Busch}(1986)}]{Busch1986Unsharp}%
  \BibitemOpen
  \bibfield  {author} {\bibinfo {author} {\bibfnamefont {P.}~\bibnamefont
  {Busch}},\ }\bibfield  {title} {\bibinfo {title} {Unsharp reality and joint
  measurements for spin observables},\ }\href
  {https://doi.org/10.1103/PhysRevD.33.2253} {\bibfield  {journal} {\bibinfo
  {journal} {Phys. Rev. D}\ }\textbf {\bibinfo {volume} {33}},\ \bibinfo
  {pages} {2253} (\bibinfo {year} {1986})}\BibitemShut {NoStop}%
\bibitem [{\citenamefont {Busch}\ and\ \citenamefont
  {Heinosaari}(2008)}]{Busch2008Approximate}%
  \BibitemOpen
  \bibfield  {author} {\bibinfo {author} {\bibfnamefont {P.}~\bibnamefont
  {Busch}}\ and\ \bibinfo {author} {\bibfnamefont {T.}~\bibnamefont
  {Heinosaari}},\ }\bibfield  {title} {\bibinfo {title} {Approximate joint
  measurements of qubit observables},\ }\href
  {https://doi.org/10.26421/QIC8.8-9-9} {\bibfield  {journal} {\bibinfo
  {journal} {Quantum Inf. Comput.}\ }\textbf {\bibinfo {volume} {8}},\ \bibinfo
  {pages} {797} (\bibinfo {year} {2008})}\BibitemShut {NoStop}%
\bibitem [{\citenamefont {Struchalin}\ \emph {et~al.}(2021)\citenamefont
  {Struchalin}, \citenamefont {Zagorovskii}, \citenamefont {Kovlakov},
  \citenamefont {Straupe},\ and\ \citenamefont
  {Kulik}}]{Struchalin2021ExperimentalShadows}%
  \BibitemOpen
  \bibfield  {author} {\bibinfo {author} {\bibfnamefont {G.~I.}\ \bibnamefont
  {Struchalin}}, \bibinfo {author} {\bibfnamefont {Y.~A.}\ \bibnamefont
  {Zagorovskii}}, \bibinfo {author} {\bibfnamefont {E.~V.}\ \bibnamefont
  {Kovlakov}}, \bibinfo {author} {\bibfnamefont {S.~S.}\ \bibnamefont
  {Straupe}},\ and\ \bibinfo {author} {\bibfnamefont {S.~P.}\ \bibnamefont
  {Kulik}},\ }\bibfield  {title} {\bibinfo {title} {Experimental estimation of
  quantum state properties from classical shadows},\ }\href
  {https://doi.org/10.1103/PRXQuantum.2.010307} {\bibfield  {journal} {\bibinfo
   {journal} {PRX Quantum}\ }\textbf {\bibinfo {volume} {2}},\ \bibinfo {pages}
  {010307} (\bibinfo {year} {2021})}\BibitemShut {NoStop}%
\bibitem [{\citenamefont {Zhang}\ \emph {et~al.}(2021)\citenamefont {Zhang},
  \citenamefont {Sun}, \citenamefont {Fang}, \citenamefont {Zhang},
  \citenamefont {Yuan},\ and\ \citenamefont
  {Lu}}]{Zhang2021ExperimentalShadows}%
  \BibitemOpen
  \bibfield  {author} {\bibinfo {author} {\bibfnamefont {T.}~\bibnamefont
  {Zhang}}, \bibinfo {author} {\bibfnamefont {J.}~\bibnamefont {Sun}}, \bibinfo
  {author} {\bibfnamefont {X.-X.}\ \bibnamefont {Fang}}, \bibinfo {author}
  {\bibfnamefont {X.-M.}\ \bibnamefont {Zhang}}, \bibinfo {author}
  {\bibfnamefont {X.}~\bibnamefont {Yuan}},\ and\ \bibinfo {author}
  {\bibfnamefont {H.}~\bibnamefont {Lu}},\ }\bibfield  {title} {\bibinfo
  {title} {Experimental quantum state measurement with classical shadows},\
  }\href {https://doi.org/10.1103/PhysRevLett.127.200501} {\bibfield  {journal}
  {\bibinfo  {journal} {Phys. Rev. Lett.}\ }\textbf {\bibinfo {volume} {127}},\
  \bibinfo {pages} {200501} (\bibinfo {year} {2021})}\BibitemShut {NoStop}%
\bibitem [{\citenamefont {Escudero}\ \emph {et~al.}(2023)\citenamefont
  {Escudero}, \citenamefont {Fern{\'a}ndez-Fern{\'a}ndez}, \citenamefont
  {Jaum{\`a}}, \citenamefont {Pe{\~n}as},\ and\ \citenamefont
  {Pereira}}]{Escudero2023HardwareEntangled}%
  \BibitemOpen
  \bibfield  {author} {\bibinfo {author} {\bibfnamefont {F.}~\bibnamefont
  {Escudero}}, \bibinfo {author} {\bibfnamefont {D.}~\bibnamefont
  {Fern{\'a}ndez-Fern{\'a}ndez}}, \bibinfo {author} {\bibfnamefont
  {G.}~\bibnamefont {Jaum{\`a}}}, \bibinfo {author} {\bibfnamefont {G.~F.}\
  \bibnamefont {Pe{\~n}as}},\ and\ \bibinfo {author} {\bibfnamefont
  {L.}~\bibnamefont {Pereira}},\ }\bibfield  {title} {\bibinfo {title}
  {Hardware-efficient entangled measurements for variational quantum
  algorithms},\ }\href {https://doi.org/10.1103/PhysRevApplied.20.034044}
  {\bibfield  {journal} {\bibinfo  {journal} {Phys. Rev. Applied}\ }\textbf
  {\bibinfo {volume} {20}},\ \bibinfo {pages} {034044} (\bibinfo {year}
  {2023})}\BibitemShut {NoStop}%
\bibitem [{\citenamefont {Klappenecker}\ and\ \citenamefont
  {R{\"o}tteler}(2005)}]{Klappenecker2005Designs}%
  \BibitemOpen
  \bibfield  {author} {\bibinfo {author} {\bibfnamefont {A.}~\bibnamefont
  {Klappenecker}}\ and\ \bibinfo {author} {\bibfnamefont {M.}~\bibnamefont
  {R{\"o}tteler}},\ }\bibfield  {title} {\bibinfo {title} {Mutually unbiased
  bases are complex projective 2-designs},\ }in\ \href
  {https://doi.org/10.1109/ISIT.2005.1523643} {\emph {\bibinfo {booktitle}
  {Proceedings of the 2005 IEEE International Symposium on Information
  Theory}}}\ (\bibinfo  {publisher} {IEEE},\ \bibinfo {year} {2005})\ pp.\
  \bibinfo {pages} {1740--1744}\BibitemShut {NoStop}%
\end{thebibliography}%
